\documentclass[symmetry,article,submit,moreauthors,pdftex]{mdpi} 

\firstpage{1} 
\pubvolume{1}
\issuenum{1}
\articlenumber{0}
\pubyear{2021}
\copyrightyear{2020}
\datereceived{} 
\dateaccepted{} 
\datepublished{} 
\hreflink{https://doi.org/} 

\usepackage{mathtools}
\usepackage{cancel}
\usepackage{amsfonts} 
\usepackage{amsmath} 
\usepackage{amssymb} 
\usepackage{bm} 
\usepackage{ragged2e} 
\usepackage{relsize} 
\usepackage{graphicx} 
\usepackage{stmaryrd} 
\usepackage{bbold}
\usepackage{amstext}
\usepackage[english]{babel}
\usepackage{helvet}
\usepackage{hyperref}

\renewcommand{\Re}{\operatorname{Re}}

\newcommand{\h}[1]{\hat{#1}}

\newcommand{\beq}{\begin{equation}}
\newcommand{\eeq}{\end{equation}}
\newcommand{\ga}{\alpha}
\newcommand{\gb}{\beta}
\newcommand{\gr}{\rho}
\newcommand{\gz}{\zeta}
\newcommand{\gga}{\gamma}
\newcommand{\chf}{confluent hypergeometric functions}
\newcommand{\che}{confluent hypergeometric equation}
\newcommand{\wta}{\widetilde{M}}
\newcommand{\wt}{\widetilde{M}(a,b,\zeta)}
\newcommand{\mbi}{\mathbb{Z}}

\Title{Continuum energy eigenstates via the factorization method}

\TitleCitation{Continuum energy eigenstates via the factorization method}

\Author{James K. Freericks $^{1}$\orcidA{}
and  W. N. Mathews Jr. $^{1\dagger}$\orcidB{}
}

\AuthorNames{James K. Freericks, Mark A. Esrick, Wesley N. Mathews Jr.}

\AuthorCitation{Freericks, J. K., Esrick, M. A., Mathews, W. N. Jr.}

\address{%
$^{1}$ \quad Department of Physics, Georgetown University, 37th and O Sts. NW, Washington, DC 20057 USA; james.freericks@georgetown.edu}

\corres{Correspondence: james.freericks@georgetown.edu
}

\firstnote{ Deceased} 

\abstract{The factorization method was introduced by Schr\"odinger in 1940. Its use in bound-state problems is widely known, including in supersymmetric quantum mechanics; one can create a factorization chain, which simultaneously solves a sequence of auxiliary Hamiltonians that share common eigenvalues with their adjacent Hamiltonians in the chain, except for the lowest eigenvalue. In this work, we generalize the factorization method to continuum energy eigenstates. Here, one does not generically have a factorization chain\;---\;instead all energies are solved using a ``single-shot factorization,'' enabled by writing the superpotential in a form that includes the logarithmic derivative of a confluent hypergeometric function. 
The single-shot factorization approach is an alternative to the conventional method of ``deriving a differential equation and looking up its solution,'' but it does require some working knowledge of confluent hypergeometric functions. This can also be viewed as a  method for solving the Ricatti equation needed to construct the superpotential.}


\begin{document}
\section{Introduction} 

Most single-particle quantum mechanics employs the coordinate representation of the Schr\"odinger equation and presents the solution for energy eigenvalue problems by using the Frobenius method \cite{Riley-etal.,McQuarrie,Boas,Arfken-etal.,Frobenius} (generalized power series solution of differential equations). This method does not work well with continuum problems, and so few textbooks discuss continuum solutions other than the free particle (in one, two, and three dimensions) and the linear potential in one dimension.

An alternative way to approach energy eigenvalue problems is via the factorization method. This is used in, for example, the textbooks by Green~\cite{Green} and Ohanian~\cite{Ohanian}. A somewhat closely related method using factorization, but emphasizing the ladder operators, is used  by Binney and Skinner~\cite{Binney-Skinner}. The factorization method was introduced by Schr\"odinger in 1940 to generalize the operator method for the simple harmonic oscillator to all other analytically solvable quantum problems~\cite{Schroedinger1,Schroedinger2,Schroedinger3}. In his original work, he argued that the method is only valid for bound states and since then, others have concurred with this conclusion~\cite{Ohanian,Dong}. Some state instead that continuum problems can be treated, but provide no details~\cite{Green,Cooper}. The textbook by de Lange and Raab~\cite{deLange-Raab} uses a ladder (or shift)  operator method (similar to, but different from the factorization method) to relate one continuum solution for the Coulomb problem to other solutions with lower values of the  angular momentum quantum number, and shows how to solve some of the free-particle problems. Rau~\cite{Rau} considered one-dimensional free continuum problems within the context of supersymmetric quantum mechanics; his methodology solves for the particle in a box and then takes the limit as the box becomes infinitely wide. Such a method is not easily generalizable to other potentials, although Rau does also discuss angular momentum solutions, which are similar. To the best of our knowledge, the other discussions of the factorization method in the literature, of which we note a small subset, \cite{Infeld,Hull-Infeld,Infeld-Hull,Adrianov-etal.1,Mielnik,Adrianov-etal.2,Alves-Filho,Marquez-etal.,Rosas-Ortiz1,Rosas-Ortiz2,Spiridonov,Dong-etal.,Mielnik-Rosas-Ortiz,Arcos-Ollala-etal.,Devi-Singh,Jacoby-etal.,Rushka-Freericks,Lyu-etal.}  restrict their attention to bound-state problems. Finally, it is also well known that one can employ these factorization methods to show that the reflection and transmission coefficients for adjacent Hamiltonians in the factorization chain are related to each other via the asymptotic behavior of the corresponding superpotentials~\cite{Cooper}.

We employ the factorization method to obtain a complete solution for all of the continuum problems that are solved with confluent hypergeometric functions:  the free particle problem in one, two, and three dimensions; the one-dimensional linear potential; hydrogen; and the one-dimensional (or Cartesian) Morse potential.  We find that all of these problems can be solved using the factorization method, but, unlike the bound state problems, which have an underlying factorization chain associated with them, the continuum solutions are solved with a ``single-shot factorization'' method\;---\;no factorization chain is possible for the general case\;---\;but the one-dimensional free particle has a special property (equally spaced nodes) that allows for a factorization chain, as we describe below. Single-shot factorization provides an alternative approach to the standard differential equations methodology. However, it does require using some subtle properties of confluent hypergeometric functions that we illustrate as we solve these different problems below.

Note that the origins of single-shot factorization lie in Schr\"odinger's original work~\cite{Schroedinger1,Schroedinger2,Schroedinger3}. There, when he solves for the particle in a box, he makes the statement that one must choose the wavevector such that it yields the highest possible energy for the ground state. This statement has reappeared in various texts~\cite{Green,Ohanian}, but there is no physical principle why this should be employed. Instead, the wavevector should be chosen so that it places nodes of the wavefunction at the boundary of the box (and this condition requires the superpotential to diverge, yielding the same result as the maximal energy condition for the particle in a box). Then, when one moves to a finite square-well potential, the condition changes from having a node at the discontinuity in the square-well, to requiring the superpotential to be continuous~\cite{Jacoby-etal.}, which then determines not just the ground state, but all bound states. In this work, we illustrate how such an approach can be generalized to the continuum energy-eigenstate problems that are solved with confluent hypergeometric functions. We note that this same technique, coupled with an appropriate boundary condition, can also be employed to solve bound state problems via factorization, without creating a factorization chain.

\section{Heuristic example of the method for the Coulomb problem}

In this section, we provide a more heuristic approach to illustrate how the single-shot factorization method works for continuum problems by solving the Coulomb problem for hydrogen. We do not present this work as a general approach, but instead, illustrate the technical manipulations needed to make the calculation work. After completing this derivation, the remainder of this work develops the general approach (which is a bit different from this heuristic approach) and then applies it to a number of different problems (including revisiting this problem).

To start, we note that the hydrogen problem is separable, so we can construct energy eigenstates as tensor products of radial states with angular momentum states, so $|\psi_{klm}\rangle=|\psi_{kl}\rangle\otimes|lm\rangle$, with $\hat{\vec{L}}^{\,2}|lm\rangle=\hbar^2 l(l+1)|lm\rangle$ and $\hat{L}_z|lm\rangle=\hbar m|lm\rangle$. Then, the radial ket $|\psi_{kl}\rangle$ satisfies
\begin{equation}
    \hat{\mathcal{H}}_l|\psi_{kl}\rangle=\left (\frac{\hat{p}_r^2}{2M}+\frac{\hbar^2l(l+1)}{2M\hat{r}^2}-\frac{e^2}{\hat{r}}\right )|\psi_{kl}\rangle=E_{kl}|\psi_{kl}\rangle.
\end{equation}
Here $\hat{p}_r$ is the radial momentum, given by
\begin{equation}
    \hat{p}_r=\frac{1}{\hat{r}}(\hat{\vec{r}}\cdot\hat{\vec{p}}-i\hbar),
\end{equation}
$M$ is the reduced mass of hydrogen, and $e$ is the magnitude of the charge of the proton and the electron.
In the factorization method approach, we factorize the Hamiltonian into the form
\begin{equation}
    \hat{\mathcal{H}}_l=\hat{A}_{kl}^\dagger\hat{A}_{kl}^{\phantom{\dagger}}+E_{kl},
\end{equation}
where, for the continuum, the energies $E_{kl}$ are all greater than zero. The index $k$ will be a continuous index, which labels the different continuum energy eigenstates. The ladder operators are represented in terms of a superpotential $W$ via
\begin{equation}
    \hat{A}_{kl}^{\phantom{\dagger}}=\frac{1}{\sqrt{2M}}\big (\hat{p}_r-i\hbar kW_{kl}(k\hat{r})\big ),
\end{equation}
where $k$ is a wavenumber, with dimensions of inverse length. Note that the factorization holds for all values of the parameter $k$, which is one of the hallmarks for how single-shot factorization works.

As a heuristic illustration, we start with a simple ansatz for the superpotential based on the factorization method for the bound states. There, the superpotential includes just a constant term and a term that is inversely proportional to $\hat{r}$. Using such a superpotential only allows one to find bound states with $E<0$. For the continuum, we add in a term proportional to the logarithmic derivative of the confluent hypergeometric function via
\begin{equation}
    W_{kl}(k\hat{r})=-\frac{\alpha}{k\hat{r}}+i\beta-\frac{1}{k}\frac{M'(a,b,2ik\hat{r})}{M(a,b,2ik\hat{r})},
\end{equation}
where $\alpha$, $\beta$, $a$, and $b$ are constants we need to determine and the prime is a derivative with respect to $\hat{r}$. The confluent hypergeometric function is used here, because it has numerous recurrence relations that are employed to allow it to factorize the hydrogen Hamiltonian with definite angular momentum.

The first step is to use the derivative relation given by
\begin{equation}
    \frac{d}{dr}M(a,b,2ikr)=2ik\frac{a}{b}M(a+1,b+1,2ikr),
\end{equation}
from the digital mathematical library of functions (DLMF) DLMF~13.3.5 \cite{DLMF}, which then leads to
\begin{equation}
    W_{kl}(k\hat{r})=-\frac{\alpha}{k\hat{r}}+i\beta-\frac{2ia}{b}\frac{M(a+1,b+1,2ik\hat{r})}{M(a,b,2ik\hat{r})}.
\end{equation}
The superpotential is required to be real (otherwise the Hamiltonian will have linear terms in radial momentum, when we form $\hat{A}^\dagger\hat{A}$). We use the Kummer transformation (DLMF 13.2.29)
\begin{equation}
    M(a,b,z)=e^zM(b-a,b,-z)
\end{equation}
to help establish this reality. If $\alpha$, $\beta$, and $b$ are real, and $\text{Re}(a)=\tfrac{1}{2}b$, then we find
\begin{align}
    \text{Im}\big (W_{kl}(kr)\big )&=\beta-\frac{2}{b}\text{Re}\left (\frac{aM(a+1,b+1,2ikr)}{M(a,b,2ikr)}\right )\nonumber\\
    &=\beta-\frac{1}{b}\text{Re}\left (\frac{aM(a+1,b+1,2ikr)}{M(a,b,2ikr)}+\frac{a^*M(a^*+1,b+1,-2ikr)}{M(a^*,b,-2ikr)}\right )\nonumber\\
    &=\beta-\frac{1}{b}\text{Re}\left (\frac{aM(a+1,b+1,2ikr)}{M(a,b,2ikr)}+\frac{a^*M(b-a+1,b+1,-2ikr)}{M(b-a,b,-2ikr)}\right )\nonumber\\
    &=\beta-\frac{1}{b}\text{Re}\left (\frac{aM(a+1,b+1,2ikr)+a^*M(a,b+1,2ikr)}{M(a,b,2ikr)}\right )\nonumber\\
    &=\beta-1,
\end{align}
because $a^*=b-a$ and because we used the identity (DMLF 13.3.3 with $b\to b+1$)
\begin{equation}
    (a-b)M(a,b+1,2ikr)+bM(a,b,2ikr)=aM(a+1,b+1,2ikr),\label{eq:ident0}
\end{equation}
which becomes
\begin{equation}
    aM(a+1,b+1,2ikr)+a^*M(a,b+1,2ikr)=bM(a,b,2ikr),
    \label{eq:ident1}
\end{equation}
because $a^*=b-a$.
Hence, we must choose $\beta=1$ to have a real superpotential. It is the Kummer relation, and recurrence relations shown below, that allow the logarithmic derivative of the confluent hypergeometric function to be part of the superpotential for hydrogen.

Our next step is to compute the real part of the superpotential. This uses the Kummer transformation again. It satisfies
\begin{align}
    &\text{Re}\big (W_{kl}(kr)\big)=-\frac{\alpha}{kr}+\frac{2}{b}\text{Im}\left (\frac{aM(a+1,b+1,2ikr)}{M(a,b,2ikr)}\right )\nonumber\\
    &~~~~~~~~~~=-\frac{\alpha}{kr}-\frac{i}{b}\left (\frac{aM(a+1,b+1,2ikr)}{M(a,b,2ikr)}-\frac{a^*M(b-a+1,b+1,-2ikr)}{M(b-a,b,-2ikr)}\right )\nonumber\\
    &~~~~~~~~~~=-\frac{\alpha}{kr}-\frac{i}{b}\left (\frac{aM(a+1,b+1,2ikr)-a^*M(a,b+1,2ikr)}{M(a,b,2ikr)}\right )\nonumber\\
    &~~~~~~~~~~=-\frac{\alpha}{kr}-\frac{i}{b}\left (\frac{2aM(a+1,b+1,2ikr)}{M(a,b,2ikr)}-b\right ),
\end{align}
after using Eq.~(\ref{eq:ident1}) again. Note that the term within the  parenthesis is purely imaginary, so that the final result is a real number.

The next step is to compute $\hat{A}_{kl}^\dagger\hat{A}_{kl}^{\phantom{\dagger}}$. We find
\begin{align}
    \hat{A}_{kl}^\dagger\hat{A}_{kl}^{\phantom{\dagger}}&=\frac{1}{2M}\left (\hat{p}_r-i\hbar\left (\frac{\alpha}{\hat{r}}-ik+\frac{2ika}{b}\frac{M(a+1,b+1,2ik\hat{r})}{M(a,b,2ik\hat{r})}\right )\right )\nonumber\\
    &~~~~~~~~\times \left (\hat{p}_r+i\hbar\left (\frac{\alpha}{\hat{r}}-ik+\frac{2ika}{b}\frac{M(a+1,b+1,2ik\hat{r})}{M(a,b,2ik\hat{r})}\right )\right )\nonumber\\
    &=\frac{\hat{p}_r^2}{2M}+\frac{i\hbar}{2M}\left [\hat{p}_r,\frac{\alpha}{\hat{r}}+\frac{2ika}{b}\frac{M(a+1,b+1,2ik\hat{r})}{M(a,b,2ik\hat{r})}\right ]\nonumber\\
    &+\frac{\hbar^2}{2M}\Bigg (\frac{\alpha^2}{\hat{r}^2}-\frac{2ik\alpha}{\hat{r}}-k^2+\frac{4ik\alpha a}{\hat{r}b}\frac{M(a+1,b+1,2ik\hat{r})}{M(a,b,2ik\hat{r})}\nonumber\\
    &~~~~~~~~~~~~~+\frac{4k^2a}{b}\frac{M(a+1,b+1,2ik\hat{r})}{M(a,b,2ik\hat{r})}-\frac{4k^2a^2}{b^2}\frac{M^2(a+1,b+1,2ik\hat{r})}{M^2(a,b,2ik\hat{r})}\Bigg ).
    \label{eq:hydrogen1}
\end{align}
Now there is some significant algebra to simplify this result. It is all straightforward, but tedious. First we compute the commutators. We have
\begin{equation}
    \left [\hat{p}_r,\frac{\alpha}{\hat{r}}\right ]=\frac{i\hbar}{\hat{r}^2},
\end{equation}
and
\begin{equation}
    \left [\hat{p}_r,\frac{M(a+1,b+1,2ik\hat{r})}{M(a,b,2ik\hat{r})}\right ]=2\hbar k\left (\frac{a+1}{b+1}\frac{M(a+2,b+2,2ik\hat{r})}{M(a,b,2ik\hat{r})}-\frac{a}{b}\frac{M^2(a+1,b+1,2ik\hat{r})}{M^2(a,b,2ik\hat{r})}\right ).
\end{equation}
When we introduce the results for the commutator into Eq.~(\ref{eq:hydrogen1}), we immediately see that the terms involving the squares of the confluent hypergeometric functions in the numerator and the denominator exactly cancel. Such a cancellation always occurs when the superpotential is written as the logarithmic derivative of function (for one value of the sign). The remaining terms become
\begin{align}
    \hat{A}_{kl}^\dagger\hat{A}_{kl}^{\phantom{\dagger}}&=\frac{\hat{p}_r^2}{2M}+\frac{\hbar^2\alpha(\alpha-1)}{2M\hat{r}^2}-\frac{i\hbar^2 k\alpha}{M\hat{r}}-\frac{\hbar^2 k^2}{2M}-\frac{2\hbar^2k^2}{M}\frac{a(a+1)}{b(b+1)}\frac{M(a+2,b+2,2ik\hat{r})}{M(a,b,2ik\hat{r})}\nonumber\\
    &+\frac{2i\hbar^2k\alpha}{M\hat{r}}\frac{a}{b}\frac{M(a+1,b+1,2ik\hat{r})}{M(a,b,2ik\hat{r})}+\frac{2\hbar^2k^2}{M}\frac{a}{b}\frac{M(a+1,b+1,2ik\hat{r})}{M(a,b,2ik\hat{r})}.
\end{align}
Our goal now is to use recurrence relations of the confluent hypergeometric function to change all of the numerators into the function in the denominator, which allows us to get rid of all of the confluent hypergeometric functions from the expression. This takes a number of steps to achieve.
First, we use Eq.~(\ref{eq:ident0}), with $a\to a+1$ and $b\to b+1$ to remove the term with indices $a+2$ and $b+2$. This yields
\begin{align}
    \hat{A}_{kl}^\dagger\hat{A}_{kl}^{\phantom{\dagger}}&=\frac{\hat{p}_r^2}{2M}+\frac{\hbar^2\alpha(\alpha-1)}{2M\hat{r}^2}-\frac{i\hbar^2 k\alpha}{M\hat{r}}-\frac{\hbar^2 k^2}{2M}-\frac{2\hbar^2k^2}{M}\frac{|a|^2}{b(b+1)}\frac{M(a+1,b+2,2ik\hat{r})}{M(a,b,2ik\hat{r})}\nonumber\\
    &+\frac{2i\hbar^2k\alpha}{M\hat{r}}\frac{a}{b}\frac{M(a+1,b+1,2ik\hat{r})}{M(a,b,2ik\hat{r})},
\end{align}
because the last term in the previous equation cancels after using the identity.

Now, we need a new identity, which is (DLMF 13.3.4 with $a\to a+1$ and $b\to b+1$)
\begin{equation}
    \frac{b+1}{2ikr}M(a+1,b+1,2ikr)-\frac{b+1}{2ikr}M(a,b+1,2ikr)=M(a+1,b+2,2ikr),
\end{equation}
and we use it to remove the $M(a+1,b+2,2ik\hat{r})$ term. This gives us
\begin{align}
    \hat{A}_{kl}^\dagger\hat{A}_{kl}^{\phantom{\dagger}}&=\frac{\hat{p}_r^2}{2M}+\frac{\hbar^2\alpha(\alpha-1)}{2M\hat{r}^2}-\frac{i\hbar^2 k\alpha}{M\hat{r}}-\frac{\hbar^2 k^2}{2M}\nonumber\\
    &+\frac{i\hbar^2k}{M\hat{r}}\frac{a}{b}\frac{a^*M(a,b+1,2ik\hat{r})+(2\alpha-a^*)M(a+1,b+1,2ik\hat{r})}{M(a,b,2ik\hat{r})}.
\end{align}
We have reached our final step, where we use the identity in Eq.~(\ref{eq:ident0}) again. Here, we note that $(a-b)=-a^*$, and if we choose $2\alpha=b$, then $2\alpha-a^*=a$, and the numerator becomes $bM(a,b,2ik\hat{r})$, so all confluent hypergeometric function terms cancel and we are left with
\begin{equation}
    \hat{A}_{kl}^\dagger\hat{A}_{kl}^{\phantom{\dagger}}=\frac{\hat{p}_r^2}{2M}+\frac{\hbar^2\alpha(\alpha-1)}{2M\hat{r}^2}-\frac{i\hbar^2 k(\alpha-a)}{M\hat{r}}-\frac{\hbar^2 k^2}{2M}.
\end{equation}
We must then choose $\alpha=l+1$, $a=l+1+\tfrac{i}{k\tilde{a}_0}$, and $b=2l+2$; here, $\tilde{a}_0=\tfrac{\hbar^2}{me^2}$ is the reduced Bohr radius). The factorization becomes
\begin{equation}
    \hat{A}_{kl}^\dagger\hat{A}_{kl}^{\phantom{\dagger}}=\frac{\hat{p}_r^2}{2M}+\frac{\hbar^2l(l+1)}{2M\hat{r}^2}-\frac{e^2}{\hat{r}}-\frac{\hbar^2 k^2}{2M},
\end{equation}
or
\begin{equation}
    \hat{\mathcal{H}}_l=\hat{A}_{kl}^\dagger\hat{A}_{kl}^{\phantom{\dagger}}+E_k,~~~\text{with}~~~E_k=\frac{\hbar^2 k^2}{2M}.
\end{equation}
One can see that this factorization holds for all $k$, so we find the subsidiary condition, given by $\hat{A}_{kl}|\psi_{kl}\rangle=0$, determines the continuum energy eigenstate for hydrogen, with energy $E_k$. The lowering operator is given by
\begin{equation}
    \hat{A}_{kl}^{\phantom{\dagger}}=\frac{1}{\sqrt{2M}}\left (\hat{p}_r+i\hbar\left (\frac{l+1}{\hat{r}}-ik+\left (ik-\frac{1}{(l+1)a_0}\right )\frac{M\left (l+2+\tfrac{i}{ka_0},2l+3,2ik\hat{r}\right )}{M\left (l+1+\tfrac{i}{ka_0},2l+2,2ik\hat{r}\right )}\right )\right ).
\end{equation}
The wavefunction is found from the subsidiary condition, by multiplying from the left with the $\langle r|$ bra, which is an eigenstate of $\hat{r}$ with eigenvalue $r$. We find the wavefunction $\psi_{kl}(r)=\langle r|\psi_{kl}\rangle$ satisfies the following differential equation:
\begin{equation}
    \frac{1}{r}\frac{d}{dr}\big (r\psi_{kl}(r)\big)=\frac{l+1}{r}-ik+\frac{d}{dr}\ln M\left (l+1+\tfrac{i}{ka_0},2l+2,2ikr\right ),
\end{equation}
because the radial momentum operator is represented by $-i\hbar\left (\tfrac{d}{dr}+\tfrac{1}{r}\right )$ in the position representation. Solving for the (unnormalized) wavefunction yields
\begin{equation}
    \psi_{kl}(r)=r^le^{ikr}M\left (l+1+\tfrac{i}{ka_0},2l+2,2ikr\right ).
\end{equation}
The requirement for the wavefunction to be a valid energy eigenstate is that the wavefunction is bounded everywhere, which this wavefunction satisfies. Note that the scattering wavefunction satisfies different boundary conditions and actually diverges at the origin. Here, we solved for the energy eigenstate, which has a wavefunction that is finite everywhere.

There are two more points to discuss about this calculation. First, when we chose the values for $\alpha$, $a$, and $b$, we chose one of the two possible solutions. The other choice was $\alpha=-l$, $a=-l+\tfrac{i}{ka_0}$, and $b=-2l$. But, if we chose those values, then the wavefunction would have behaved like $r^{-l-1}$ as $r\to 0$ which goes to infinity and is not an acceptable solution. Second, the wavefunction needs to be normalized via a delta function normalization. This can be done in many different ways (energy normalized, momentum normalized, etc.) and is discussed thoroughly in textbooks, so we do not discuss it further here.

In this section, we provided a heuristic approach to how one uses single-shot factorization to solve for the hydrogen problem. In this case, we used a logarithmic derivative of the confluent hypergeometric function plus simple functions (constants plus powers) for the superpotential, and then forced it to work. In the remainder of this paper, we describe the general way to approach this problem, which separates the conditions of showing one has an eigestate, versus requiring reality of the superpotential and boundedness of the wavefunction, and then employ that method to solve all of the Hamiltonians that have confluent hypergeometric functions as the solutions of their continuum energy eigenstates.

\section{Development of the General Form of the Superpotential}

We consider a particle of mass $M$ (which can be a reduced mass) moving in a potential given by $V(q)$, where we use the general symbol $q$ for the position coordinate (which is $x$ in one dimensions $\rho=\sqrt{x^2+y^2}$ in two dimensions, and $r=\sqrt{x^2+y^2+z^2}$ in three dimensions). When we work with circular (or spherical) symmetric problems in two (and three) dimensions, the effective potential includes the original potential plus the centripetal potential. The problems in higher dimensions are separable, so we write the state as $|\psi_\rho\rangle\otimes|m\rangle$ in two dimensions and $|\psi_r\rangle\otimes|lm\rangle$ in three dimensions. Here $\hat{L}_z|m\rangle=\hbar m|m\rangle$, while $\hat{\vec{L}}^{\,2}|lm\rangle=\hbar^2l(l+1)|lm\rangle$ and $\hat{L}_z|lm\rangle=\hbar m|lm\rangle$. When the Hamiltonian acts on this state, it can be represented in terms of a constant angular momentum Hamiltonian, which we denote as $\hat{\mathcal{H}}_m$ in two dimensions and $\hat{\mathcal{H}}_l$ in three dimensions. Finally, the kinetic energy has now been divided into the radial motion (determined by the corresponding radial momentum $\hat{p}_q$ in higher dimensions) and the angular contribution, which is absorbed into the effective potential.
The effective potential then takes the following form:
\beq \label{Eq:VT}
V_{\text{eff}} (\h{x}) = 
\begin{cases}
V(\h{x}), \; &\text{for Cartesian coordinates} \\
V(\h{\gr}) + \frac{\hbar^2\left ( m^2 - \tfrac{1}{4}\right )}{2 M \h{\gr}^2}, \; &\text{for  plane polar coordinates} \\
V(\h{r}) + \frac{\hbar^2l (l+1)}{2M\h{r}^2}, \; &\text{for spherical coordinates},
\end{cases}
\eeq
with the Hamiltonian given by 
\begin{equation}
    \hat{\mathcal{H}}=\frac{\hat{p}_q^2}{2M}+V_{\text{eff}}(\hat{q}).\label{eq:H}
\end{equation}

We wish to factorize the Hamiltonian according to
\beq \label{Eq:Factorization}
\mathcal{\h{H}} = \h{A}_k^{\dagger} \h{A}_k  + E_k,
\eeq
where $E_k$ is the energy eigenvalue in our construction,  and the lowering operator is given by
\beq \label{Eq:A 1}
\h{A}_k = \frac{1}{\sqrt{2M}}\left[\h{p}_q - i \hbar k W_k(k\h{q})\right].
\eeq
Here, $W_k$ is known as the superpotential \cite{Cooper}.  The quantity $k$ is a wavenumber introduced so that the argument of the superpotential and the superpotential itself are dimensionless.   Equation \eqref{Eq:Factorization}, and the Schr\"{o}dinger eigenvalue equation,
\beq
\hat{\mathcal{H}}|\psi_k\rangle = E_k |\psi_k\rangle,
\eeq
together imply the subsidiary condition associated with $\h{A}_k$, 
\beq
\h{A}_k|\psi_k\rangle = 0\;,
\eeq
as the defining relation for the energy eigenstate with energy $E_k$.
This follows simply because the factorized form of the Hamiltonian is a positive semidefinite form, minimized by the state that satisfies the subsidiary condition.
This subsidiary condition in turn determines the system state vectors, $|\psi_k\rangle$, and leads to 
\beq \label{Eq:Superpotential1}
W_k(kq) = - \frac{d}{d z} \left[\ln \psi_k(q)\right],
\eeq
where we have taken $z = kq + \ga$, with $\ga$ a constant.  Here,
\beq \label{Eq:Wavefunctions}
\psi_k(q) = 
\begin{cases}
\psi_k(x) \; &\text{for Cartesian coordinates} \; , \\
\sqrt{\gr} \psi_{k\rho}(\gr) \; &\text{for plane polar coordinates} \; \\
r \psi_{kr}(r) \; &\text{for spherical coordinates} \; ,
\end{cases}
\eeq
where $\psi_k (x), \psi_{k\rho}(\gr)$, and $\psi_{kr}(r)$ are the wavefunctions for Cartesian coordinates, and (the radial wavefunctions) for plane polar and spherical coordinates, respectively.

In the absence of an electromagnetic vector potential, the Hamiltonian or Hamiltonian with definite angular momentum, must have no linear terms in momentum, which requires that the superpotential be real. We will make this requirement for all the work in this paper.

Since many of the solutions of the Schr\"{o}dinger equation can be expressed in terms of products of \chf ~with other simple functions~\cite{Natanzov,Cordero-etal.,Pena-etal.}, we choose
\beq \label{Eq:Superpotential2}
W_k(kq) = -  \frac{d}{d z} \lbrace \ln\left[f(z)F(a,b,\gz)\right]\rbrace ,\;z = k q + \ga , \; \gz = \gz (z) ,
\eeq
where: $f$ is a function of $z$ to be determined; $F$ is one of the \chf, $M, U$ \cite{DLMF,AS}, or $\gz^{1-b}\widetilde{M}$, where we define $\wta$ below; $a, b$ and $\ga$ are parameters to be determined; $\gz$ is also a function of $z$ to be determined.  (We reference \cite{DLMF,AS} as DLMF and AS, respectively.)

Solving the confluent hypergeometric equation involves a number of subtleties, which are discussed in detail in Ref.~\cite{Properties}. The issue is that most differential equations that are solved by special functions, have the special functions defined to be the different linearly independent solutions. But, for hypergeometric functions (and confluent hypergeometric functions) they are first defined in terms of a power series, or related formulas, which leads to a curious result, that for some values of the parameters that enter the differential equation, the different confluent hypergeometric functions become proportional to each other. Hence, a general approach to solving the confluent differential equation requires a methodology that takes this possible occurrence into account. In other words, there is no simple way to identify the two linearly independent solutions for all $a $ and $b$. Instead one has to use a more complex procedure when solving the differential equation.
Hence, to solve the \che,
\beq
\gz \frac{d^2w}{d \gz^2} + (b - \gz) \frac{d w}{d \gz} - a w = 0\;,
\eeq
one should consider not only $M(a,b,\gz)$ and $U(a,b,\gz)$ as potential solutions, but also
\beq \label{Eq:Mtilde}
\widetilde{M}(a,b,\gz) \equiv \gz^{1-b}\;M(1+a-b,2-b,\gz)\;.
\eeq
The motivation for this definition of $\widetilde{M}(a,b,\gz)$ is that it is a very convenient and useful shorthand.  The significance of the Kummer function, $M(a,b,\gz)$, and the function $\widetilde{M}(a,b,\gz)$ is that they are, respectively, the first and second generalized or Frobenius power series solutions \cite{Riley-etal.,McQuarrie,Boas,Arfken-etal.,Frobenius} of the \che. 

We next use one of the standard identities for $F' \equiv d F/d\gz$ (13.3.20 and 13.3.27 with $n=1$ of DLMF, or 13.4.12 and 13.4.25 of AS),
\beq \label{Eq:Identity 1}
F'(a,b,\gz) = F(a,b,\gz) - \gb F(a,b+1,\gz),
\eeq
(with $\beta$ defined below) and obtain
\beq \label{Eq:Superpotential3}
W_k(kq) = g(z) + \gb \frac{d \gz}{d z} \frac{F(a,b+1,\gz)}{F(a,b,\gz)} ,
\eeq
where
\beq \label{Eq:g and f}
g(z) = - \frac{d}{d z} \ln[e^\gz f(z)]
\eeq
and
\beq \label{Eq:beta}
\gb =
\begin{cases}
\frac{b-a}{b}, &\text{when we are using}\;M(a,b,\gz) \\
1,             &\text{when we are using}\;U(a,b,\gz) \\
\tfrac{1-a}{2-b},&\text{when we are using}\;\gz^{b-1}\widetilde{M}(a,b,\gz) \equiv M(1+a-b,2-b,\gz)\;.
\end{cases}
\eeq
The exponential term in $g(z)$ arises from the $F(a,b,\zeta)$ piece of Eq.~(\ref{Eq:Identity 1}), which is then combined with the $f$ term.

As a consequence of the requirement that $W_k (kq)$ is real, we can now write
\beq \label{Eq:A 2}
\h{A}_k = \frac{1}{\sqrt{2M}}\left\{ \h{p}_q - i \hbar k \left[ g(\h{z}) + \gb \frac{d \h{\gz}}{d \h{z}} \frac{F(a,b+1,\h{\gz})}{F(a,b,\h{\gz})} \right] \right\}
\eeq
and
\beq \label{Eq:A dagger}
\h{A}_k^{\dagger} = \frac{1}{\sqrt{2M}}\left\{ \h{p}_q + i \hbar k \left[ g(\h{z}) + \gb \frac{d \h{\gz}}{d \h{z}} \frac{F(a,b+1,\h{\gz})}{F(a,b,\h{\gz})} \right] \right\}.
\eeq
Note that this may not look like we took the Hermitian conjugate properly, but we did, because we are requiring the superpotential to be real valued (and we will use this requirement to reject unacceptable solutions). We already saw this issue in the hydrogen example in the previous section, where we used the Kummer transformation to verify reality of the superpotential.

\section{Details of the Superpotential}

We now use Eqs.~\eqref{Eq:A 2} and \eqref{Eq:A dagger} to  calculate
\begin{align}
\h{A}_k^{\dagger}\h{A}_k &= \frac{1}{2M}\left\{\h{p}_q^2 + \hbar^2 k^2 \Big[ g^2(\h{z}) + 2 \gb g(\h{z}) \frac{d \h{\gz}}{d \h{z}} \frac{F(a,b+1,\h{\gz})}{F(a,b,\h{\gz})} + \right.\nonumber\\ 
&\left.+ \gb^2 \left(\frac{d \h{\gz}}{d \h{z}}\right)^2 \frac{F^2(a,b+1,\h{\gz})}{F^2(a,b,\h{\gz})}\Big] - i\hbar k \left [\h{p}_q,g(\h{z}) + \gb \frac{d \h{\gz}}{d \h{z}} \frac{F(a,b+1,\h{\gz})}{F(a,b,\h{\gz})} \right]\right\}.
\end{align}
After evaluating the commutator, and using the identity of Eq. \eqref{Eq:Identity 1} and (See 13.3.18 and 13.3.25 of DLMF, or 13.4.13 and 13.4.24 of AS with $b \to b+1$. Note also that here $\gamma$ is not Euler's constant.),
\beq
\gz F'(a,b+1,\gz) = - \gamma F(a,b+1,\gz) + \delta F(a,b,\gz),
\eeq
where 
\beq \label{Eq:gamma}
\gamma =
\begin{cases}
b,		&\text{for}\;F(a,b,\gz) = M(a,b,\gz) \\
b,      &\text{for}\;F(a,b,\gz) = U(a,b,\gz) \\
2-b,    &\text{for}\;F(a,b,\gz) = \gz^{b-1}\;\widetilde{M}(a,b,\gz)\equiv M(1+a-b,2-b,\gz)\;,
\end{cases}
\eeq
and
\beq \label{Eq:delta}
\delta =
\begin{cases}
b,		&\text{for}\;F(a,b,\gz) = M(a,b,\gz) \\
b-a,    &\text{for}\;F(a,b,\gz) = U(a,b,\gz) \\
2-b,    &\text{for}\;F(a,b,\gz) = \gz^{b-1}\;\widetilde{M}(a,b,\gz)\equiv M(1+a-b,2-b,\gz)\;,
\end{cases}
\eeq
we obtain
\begin{align} \label{Eq:Crunch time}
\h{A}_k^{\dagger}\h{A}_k &= \frac{1}{2M}\Bigg\{\h{p}_q^2 + \hbar^2 k^2 \left[g^2(\h{z}) -\frac{d g(\h{z})}{d \h{z}} - \frac{\gb \delta}{\h{\gz}} \left(\frac{d \h{\gz}}{d \h{z}}\right)^2 \right]\nonumber\\ 
&+ \gb \hbar^2 k^2  \left[2g(\h{z}) \frac{d \h{\gz}}{d \h{z}} + \left(1 + \frac{\gamma}{\h{\gz}}\right) \left(\frac{d \h{\gz}}{d \h{z}}\right)^2 - \frac{d^2 \h{\gz}}{d \h{z}^2} \right]\frac{F(a,b +1,\h{\gz})}{F(a,b,\h{\gz})} \Bigg\} .
\end{align}
Note that the terms involving the squares of the confluent hypergeometric functions have cancelled, as we alluded to in the previous section.

We have reached the point at which we must figure out how to determine $a, b, \gz$, and $g$.  The key point is that we need
\beq
\h{A}_k^{\dagger}\h{A}_k = \frac{\h{p}_q^2}{2M} + V_{\text{eff}}(\h{q}) - E_k,
\eeq
in order to have factorized the Hamiltonian.
This requires that
\begin{align} \label{Eq:Gen Cond}
&\frac{\hbar^2 k^2}{2 M} \left[g^2(z) -\frac{d g(z)}{d z} -  \frac{\gb \delta}{\gz} \left(\frac{d \gz}{d z}\right)^2 \right] + \nonumber\\ 
&+ \gb \frac{\hbar^2 k^2}{2 M} \left[2g(z) \frac{d \gz}{d z} + \left(1 + \frac{\gamma}{\gz}\right) \left(\frac{d \gz}{d z}\right)^2 - \frac{d^2 \gz}{d z^2} \right]\frac{F(a,b +1,\gz)}{F(a,b,\gz)} = \; V_{\text{eff}}(q) - E_k  .
\end{align}
The simplest, but obviously not most general, way to satisfy this requirement is to separately set
\beq \label{Eq:Cond 1}
g^2(z) -\frac{d g(z)}{d z} - \frac{\gb \delta}{\gz} \left(\frac{d \gz}{d z}\right)^2 = \frac{2M}{\hbar^2 k^2} \left[ V_{\text{eff}}(q) - E_k \right]
\eeq
and
\beq \label{Eq:g 1}
g(z) = - \frac{1}{2} \left(1+\frac{\gamma}{\gz}\right) \frac{d \gz}{d z} + \frac{1}{2} \frac{d^2 \gz/d z^2}{d \gz/d z} .
\eeq
(The question of whether other options are possible remains an open question that we will not consider further in this work.)
We substitute this equation for $g(z)$ into Eq. \eqref{Eq:Cond 1} to obtain
\begin{align}
\left(\frac{d \gz}{d z}\right)^2 &\left[1 + \frac{2(\gamma-2\gb\delta)}{\gz} + \frac{\gamma(\gamma-2)}{\gz^2} \right] + 3 \left( \frac{d^2 \gz/d z^2}{d \gz/d z}\right)^2  - 2  \frac{d^3 \gz/d z^3}{d \gz/d z} = \\ &= \frac{8M}{\hbar^2 k^2} \left[ V_{\text{eff}}(q) - E_k \right] \notag \; . 
\end{align}
We note from Eqs.~\eqref{Eq:beta},~ \eqref{Eq:gamma}, and \eqref{Eq:delta} that $\gamma-2\gb\delta = 2a-b$ and $\gamma(\gamma-2) = b(b-2)$\;.  Thus the last equation reduces to
\beq \label{Eq:Zeta}
\left(\frac{d \gz}{d z}\right)^2 \left[1 + \frac{2(2a-b)}{\gz} + \frac{b(b-2)}{\gz^2} \right] + 3 \left( \frac{d^2 \gz/d z^2}{d \gz/d z}\right)^2  - 2  \frac{d^3 \gz/d z^3}{d \gz/d z} =  \frac{8M}{\hbar^2 k^2} \left[ V_{\text{eff}}(q) - E_k \right] \; . 
\eeq
Note that this is a complex nonlinear high-order differential equation. But, we will find for all solutions we consider, the function $\zeta(z)$ is either a linear function, a power law, or an exponential. In all three of these cases, the nonlinear differential equation greatly simplifies and this allows for a direct solution of this problem.
From Eq.~\eqref{Eq:g 1}, we see that
\beq \label{Eq:g2}
g(z) = -\frac{d}{d z} \ln \left[\frac{e^{\gz/2} \gz^{\gamma/2}}{\left(d \gz/d z \right)^{1/2}}\right]\;.
\eeq
It follows from Eqs.~\eqref{Eq:g and f} and \eqref{Eq:g2} that
\beq \label{Eq:f1}
f(z) = \mathcal{C} \frac{e^{-\gz/2} \gz^{\gamma/2}}{\left(d \gz/d z \right)^{1/2}}\;,
\eeq
where $\mathcal{C}$ is a constant,

Note carefully that in Eqs.~\eqref{Eq:beta},       
 \eqref{Eq:gamma}, and \eqref{Eq:delta}, the third instance is not for \\
 ${F(a,b,\gz) = \widetilde{M}(a,b,\gz)}$, but is for $F(a,b,\gz) = \gz^{b-1}\;\widetilde{M}(a,b,\gz) \equiv M(1+a-b,2-b,\gz)$.  Thus, Eq.~\eqref{Eq:gamma}, and Eq.~\eqref{Eq:f1} allows us to make one last simplification.  We see, on the one hand, that when $F = M$ or $U$,
\beq
f(z)F(a,b,\gz) = \mathcal{C}\frac{e^{-\gz/2} \gz^{b/2}}{\left(d \gz/d z \right)^{1/2}} F(a,b,\gz)~.
\eeq
On the other hand
\beq
f(z)\gz^{b-1}\widetilde{M}(a,b,\gz) = \mathcal{C}\frac{e^{-\gz/2} \gz^{1-b/2}}{\left(d \gz/d z \right)^{1/2}} \gz^{b-1}\widetilde{M}(a,b,\gz) = \mathcal{C}\frac{e^{-\gz/2} \gz^{b/2}}{\left(d \gz/d z \right)^{1/2}}\widetilde{M}(a,b,\gz)~.
\eeq
Consequently
\beq \label{Eq:f2}
f(z)F(a,b,\gz) = h(z)~\mathfrak{F}(a,b,\gz)~,
\eeq
where
\beq \label{Eq:h1}
h(z) = \mathcal{C}\frac{e^{-\gz/2} \gz^{b/2}}{\left(d \gz/d z \right)^{1/2}}
\eeq
and $\mathfrak{F}$~is~${M,~U}$, {or}~ ${\wta}$~.  It follows that
\beq \label{Eq:Superpotential5}
W_k(kq) = -  \frac{d}{d z} \lbrace \ln\left[h(z)\mathfrak{F}(a,b,\gz)\right]\rbrace , \; z = k q + \ga , \; \gz = \gz (z)~,~\mathfrak{F}= M,~U,~ \text{or}~\wta,
\eeq
and
\beq \label{Eq:psiofq}
\psi_k(q) \propto h(z)~\mathfrak{F}(a,b,\gz)~,~z=kq+\ga,~\gz=\gz(z),~ \mathfrak{F}= M,~U,~\text{or}~\wta .
\eeq

Equations \eqref{Eq:Zeta} and \eqref{Eq:g2} together offer the simplest, though not most general, way of satisfying Eq. \eqref{Eq:Gen Cond}.  Once we have determined $a, b, h(z)$, and $\gz(z)$, using Eq.~\eqref{Eq:Gen Cond}, \eqref{Eq:g2}, \eqref{Eq:f1}, and \eqref{Eq:f2}, or Eqs.~\eqref{Eq:Zeta} and \eqref{Eq:h1}, the remaining task is to determine which of the confluent hypergeometric functions, $M(a,b,\gz), U(a,b,\gz)$, or $\wt$, to use.  As we have noted, this is determined by the behavior of the \chf\;at the boundaries of the domain we are solving the Schr\"odinger equation on and the conditions of the wavefunction that it is finite everywhere, real, and, in some cases, vanishes at the boundary.

What we have done up to this point is valid for any potential $V(q)$ that requires $M(a,b,\gz), U(a,b,\gz)$, or $\wt$.  Our results are summarized by Eqs.~ \eqref{Eq:VT}, \eqref{eq:H}, \eqref{Eq:Wavefunctions}, \eqref{Eq:Mtilde}, \eqref{Eq:Superpotential2}, \eqref{Eq:Gen Cond}, \eqref{Eq:Zeta}, \eqref{Eq:g2}, \eqref{Eq:f1}, \eqref{Eq:f2}, \eqref{Eq:h1}, \eqref{Eq:Superpotential5}, and \eqref{Eq:psiofq}.  These results apply to problems employing Cartesian, plane polar and spherical coordinates, provided $V_{\text{eff}}(q)$ is interpreted according to Eq. \eqref{Eq:VT}.  Thus we now have a general procedure for dealing with at least the simpler (i.e., those compatible with Eqs. \eqref{Eq:Zeta} and \eqref{Eq:g2}) continuum state problems that can be solved with \chf. In the remainder of this work, we show how to solve five different problems using this methodology.

But before that, we have a few comments to make. Note how the potential is determined by the function $\zeta(z)$, and the wavefunction (through the superpotential) is determined by the confluent hypergeometric function. This has thereby separated the solution of these problems into two steps: (i) first we determine the function $\zeta(z)$, which determines the potential and sets the allowed values for a number of constants appearing in the solution, and (ii) then we determine the confluent hypergeometric function by requiring that the wavefunction satisfies the appropriate properties (bounded everywhere and real). This separation into two stages is very similar to the Nikiforov-Uvarov (NU) method of solving hypergeometric equations~\cite{nu-method}. The NU method is often used for finding bound states, but we are not aware of any application to continuum states. The approach we give here is similar in spirit, but the details are quite different. However, it would be easy to believe that they might be related in some way. Finally, we will find in many cases there is more than one way to solve the problems that lead to the same final wavefunctions. In these situations, we find that the ladder operators end up being the same, even though the pathway toward finding them is different for the different approaches. This occurs because different confluent hypergeometric functions are proportional to each other for special values of their parameters. We discuss how this works in a case-by-case basis below for the different example solutions.

\section{Free Particle in One Cartesian Dimension} \label{S:1D}

For the free particle in one dimension,  the simplest choice for $\gz$ is $\gz = cz$, where $z=kx+\ga$, and $c$ and $\ga$ are constants to be determined.  Equation \eqref{Eq:Zeta} then reduces to
\beq \label{Eq:Par rels 1}
(c^2 + 4) + \frac{2(2a-b)c}{z} + \frac{b(b-2)}{z^2} = 0 ,
\eeq
with $E_k$ given by
\beq \label{Eq:Energy 1}
E_k = \frac{\hbar^2 k^2}{2m} \; .
\eeq
Since this result must hold for all values of $z$, we find that
\beq
c=\pm 2i,\;b = 2a,\;b = 0\;\text{or}\;2\;.
\eeq
We note that for $a=b=0, M(a,b,\gz)$ does not exist, and for $a=1, b=2, \tilde{M}(a,b,\gz)$ does not exist.  Consequently
\begin{align}
c = \pm 2i,\;&a=0,\;b=0\;,~\text{allows for}~\wta ~\text{and}~U~,\\
c = \pm 2i,\;&a=1,\;b=2\;,~\text{allows for}~M~\text{and}~U~.
\end{align}
Thus we have the possibilities as shown in Table I below.  The ``Comments'' are the results of the following analysis. The choices for the \chf~to use in all of the systems we consider is nuanced and covered in full elsewhere~\cite{Properties}.
\begin{table}[hbt] 
\begin{tabular}[t]{||l|l|l|l|l|l|l||}
\hline \hline
Case \#&$\;a$   &$\;b$  &$\quad c$   &\;$\mathfrak{F}\;$&\;Comments                 \\ \hline \hline
1      &$\;0\;$ &$\;0\;$&$\;\pm 2i\;$&\;$\;\wta\phantom{\Big |}$        &\;Usual Solutions          \\ \hline
2      &\;0     &\;0    &$\;\pm 2i\;$&\;$\;U\phantom{\Big |}$           &\;Imaginary Superpotential \\  \hline
3      &\;1     &$\;2\;$&$\;\pm 2i\;$&\;$\;M\phantom{\Big |}$           &\;Usual Solutions          \\ \hline
4      &\;1     &\;2    &$\;\pm 2i\;$&\;$\;U\phantom{\Big |}$           &\;Imaginary Superpotential \\ \hline \hline
\end{tabular}
\caption{Possible solutions for the free particle in one  dimension.} \label{T:Table1}
\end{table} 

For $\gz=\pm2iz$, Eq.\eqref{Eq:h1} gives
\beq \label{Eq:f 1}
h(z) = \mathcal{C} \frac{e^{\mp i z} (\pm 2iz)^{b/2}}{\sqrt{\pm2i}}\;.
\eeq
Thus the superpotential is given by
\beq \label{Eq:Superpotential4}
W_k(kx) = -  \frac{d}{d z} \Bigg\{ \ln \left[e^{\mp i z} z^{\gamma/2} \mathfrak{F}(a,2a,\pm 2iz)\right]\Bigg\} , z = kx + \ga \;.
\eeq

For case \# 1, since according to Eq.~\eqref{Eq:Mtilde},
\beq
(\pm 2iz)^{-1}\;\widetilde{M}(0,0,\pm 2iz) = M(1,2,\pm 2iz)\;,
\eeq
we have
\beq
W_k (kx) = - \frac{d}{d z} \Bigg\{ \ln \left[ e^{\mp i z} z\;M(1,2,\pm 2iz)\right ]\Bigg\}, z = kx + \ga .
\eeq
However (See 13.6.2 of DLMF and 13.6.13 of AS.),
\beq \label{Eq:SupArg}
\sin(z) = e^{\mp i z} z\;M(1,2,\pm 2iz) .
\eeq
(The $\mp$ and $\pm$ are allowed because $a,b$, and $z$ are real.)  Consequently,
\beq \label{Eq:Superpotential5a}
W_k (kx) = - \cot(z)\;, \psi_k(x) \propto \sin(z)\;, z = kx+\ga\;.
\eeq
Since the Hamiltonian commutes with the parity operator, we can find simultaneous eigenstates of the Hamiltonian and parity operators, i.e., even and odd solutions.  So, for $\ga = \pi/2$ and $\ga = 0$, we obtain
\beq \label{Eq:Wavefunction1}
\psi_{ke} (x) = \mathcal{N}_{ke} \cos (kx), \; \psi_{ko} (x) = \mathcal{N}_{ko} \sin(kx) ,
\eeq
where $e$ and $o$ stand for ``even'' and ``odd'', respectively, and where  $\mathcal{N}_{ke}$ and $\mathcal{N}_{ko}$ are the respective normalization constants.
As we have noted, the superpotential must be real and, from Eq. \eqref{Eq:Superpotential5a}, , we see that this requirement is indeed satisfied for the valid solutions.

We see that in Eqs.~\eqref{Eq:Energy 1} and \eqref{Eq:Wavefunction1} we have a complete solution for the problem, with just a single factorization.  No ladder of auxiliary Hamiltonians is needed. Note further that it is the requirement of reality of the superpotential and verifying the finiteness of the wavefunction that determines the final result that can be used as solutions.

For case \# 2,
\beq
\psi_k (x) \propto e^{\mp iz}\;U(0,0,\pm 2 i z)\;.
\eeq
According to 13.2.40 of DLMF and 13.1.29 of AS,
\beq 
U(a,b,\gz) = \gz^{1-b} U(a-b+1,2-b,\gz)\;,
\eeq
so that
\beq
U(0,0,\gz) = \gz\;U(1,2,\gz)\;,
\eeq
and since (See 13.6.4 of DLMF.)
\beq \label{Eq:U}
U(a,a+1,\gz) = \gz^{-a}\;,
\eeq
it follows that
\beq
U(0,0,\gz)=1\;.
\eeq
Consequently,
\beq
\psi_k (x) \propto e^{\mp iz}\, 
\eeq
which implies an imaginary superpotential, which is not consistent.  Thus no acceptable solution results.

For case \# 3 and case \# 4, the wavefunctions are
\beq
\psi_k (x) \propto e^{\mp i z} z\;M(1,2,\pm 2iz), z = kx + \ga .
\eeq
and
\beq
\psi_k (x) \propto e^{\mp i z}\;z\;U(1,2,\pm 2iz), z = kx + \ga .
\eeq
These are the same as for case \# 1 and case \# 2, respectively.  Thus, the results shown in Table \ref{T:Table1} follow. Note that the two valid solutions have identical superpotentials, just as we had claimed they would.

For the one-dimensional free particle problem, we actually can generate a factorization chain.  With Eq.~\eqref{Eq:Superpotential5a} in mind, we take
\beq \label{Eq:Ajk}
\h{A}_{kj} = \frac{1}{\sqrt{2M}}\; [\;\h{p}_x + i\hbar k_j \cot(\h{z})\;],\;z = kx+\alpha\;,\;j=0, 1, 2, \ldots \;.
\eeq
The subscript $j$ is necessary to show us which link on the chain we are considering; our previous work in this section has been for $j=0$.  We readily find that in general,
\beq \label{Eq:Hjk}
\h{\mathcal{H}}_{j} = \frac{\h{p}_x^2}{2M} + j(j+1) E_{k0}\csc^2 (\h{z})\;,
\eeq
\beq \label{Eq:kj}
k_j = (j+1)\;k,
\eeq
and 
\beq \label{Eq:Ejk}
E_{kj} = \frac{\hbar^2 [(j+1)k]^2}{2M}\;.
\eeq
The subsidiary condition,
\beq
\h{A}_{kj}\;|\phi_{kj} \rangle = 0\;,
\eeq
and Eqs.~\eqref{Eq:Ajk} and \eqref{Eq:kj} give
\beq
\left[\frac{d}{d x} - (j+1)k\cot{z}\right] \phi_{kj} (x) = 0\;,
\eeq
which has the solution
\beq 
\phi_{kj} (x) = \mathcal{N}_{kj}^{\;'}\;\sin^{j+1} (z)\;.
\eeq
The statevectors for the $jth$ excited state of $\h{\mathcal{H}}_0$ are given by
\beq \label{Eq:psij}
|\psi_{jk} \rangle = \h{A}_{k,0} ^{\dagger} \ldots \h{A}_{k,j-1} ^{\dagger}\;|\phi_{kj}\rangle\;,\;\text{for}\;j \ge 1\;.
\eeq
We readily obtain
\beq
\psi_{kj} (x) = \mathcal{N}_{kj}\;\sin[(j+1)z]\;,\;\text{for}\;j = 1, 2, \ldots \; .
\eeq
Then,
\beq
\psi_{kje}(x) =  \mathcal{N}_{kje}\;\cos[(j+1)kx],\; \psi_{kjo}(x) = \mathcal{N}_{kjo}\;\sin[(j+1)kx]
\eeq
are the wavefunctions for the $j$th excited state of $\h{\mathcal{H}} \equiv \h{\mathcal{H}}_0$. On the one hand, there is little difficulty in obtaining these results.  On the other hand, the effort does not get us anything new or result in any additional insight. What is interesting is that Schr\"odinger already derived these results when solving the particle in a box~\cite{Schroedinger1}, but did not equate the solution as a means for generating continuum solutions. Instead, he stated that continuum solutions were not possible with the factorization method.

We note that all of the auxiliary Hamiltonians ($j\ge 1$) contain trigonometric P\"oschl-Teller potentials, albeit repulsive ones~\cite{Flugge}.  Moreover, they all have singularities at $z$ equal to integral multiples of $\pi$.  That is, the auxiliary Hamiltonians all have singularities at the same periodic values of $z$\;.  Nevertheless, there is a continuum of solutions for each auxiliary Hamiltonian because the wave functions vanish at the singularities.  This is indeed critical for the existence of the factorization chain in the continuum. It requires the nodes to be equally spaced, so that the nodes for one $j$ are all included in the nodes for all higher $j$. This property only holds in this Hamiltonian. Note further that the factorization chain does not include all continuum energy eigenvalues. Rather, there are an infinite continuum of independent factorization chains,  required to obtain all of the solutions. In this sense, the existence of the factorization chain is not particularly helpful for these continuum problems, but it is an interesting curiosity that they do exist. They do not exist for the other solutions we consider in this work.

\section{Free Particle in Two Dimensions} \label{S:2D}

For the free particle in two-dimensions, we use plane polar coordinates and factor the effective radial Hamiltonian.  Again, the simplest choice for $\gz$ is $\gz = cz$.  Moreover, because of the form of $V_{\text{eff}}(\h{\gr})$, we also take $z = k \gr$. Equation \eqref{Eq:Zeta} then reduces to
\beq
(c^2 + 4) + \frac{2(2a-b)c}{z} + \frac{b(b-2)-(4m^2 -1)}{z^2} = 0 ,
\eeq
where we have again used Eq. \eqref{Eq:Energy 1}.  We see that we must take
\beq
c=\pm 2i, b=2a, b=1\pm2|m|~.
\eeq
Moreover, we note that for $b=1-2|m|$, we should use $\wta$ and $U$, and for $b=1+2|m|$, we should use $M$ and $U$.  Consequently
\begin{align}
c&=\pm2i, a = \tfrac{1}{2} - |m|, b = 1 - 2|m|~,~\text{allows for}~\wta ~\text{and}~U~,\\
c&=\pm2i, a = \tfrac{1}{2} + |m|, b = 1 + 2|m|~,~\text{allows for}~M~\text{and}~U~.
\end{align}

We now have to examine all of these possible cases for the superpotential and for the wavefunction. 
These possibilities as summarized in Table II.  The * after $\wta$ is to remind us that for $m=0, ~b=1$, we have $\wta=M$, so that we can simply use $\wta$ for item \#1. The comments are again the result of the following analysis.
\begin{table}[h!] 
\begin{tabular}[t]{||l|l|l|l|l|l|l||}
\hline \hline
Case \#&$\;a$                 &$\;b$       &$\quad c$   &\;$\mathfrak{F}\;$   &\;Comments  \\ \hline \hline
1      &$\;\tfrac{1}{2}-|m|\;$&$\;1-2|m|\;$&$\;\pm2i\;$ &$\;\wta^*\phantom{\Big |}$&\;Usual Solution for $m \le 0$ \\ \hline
2      &$\;\tfrac{1}{2}-|m|\;$&$\;1-2|m|\;$&$\;\pm 2i\;$&$\;\,U\phantom{\Big |}$              & \;Imaginary Superpotential \\  \hline
3      &$\;\tfrac{1}{2}+|m|\;$&$\;1+2|m|\;$&$\;\pm 2i\;$&$\;M\phantom{\Big |}$                &\;Usual Solution for $m \ge 0$ \\ \hline
4      &$\;\tfrac{1}{2}+|m|\;$&$\;1+2|m|\;$&$\;\pm 2i\;$&$\;\,U\phantom{\Big |}$              &\;Imaginary Superpotential \\ \hline \hline
\end{tabular}
\caption{Possible solutions for the free particle in plane polar coordinates.} \label{T:Table2}
\end{table} 

From Eq.~\eqref{Eq:h1}, with $\gz(z)=(\pm)_{\gz}\;2iz$ and $b = 1 (\pm)_b 2|m|$, we have
\beq
h(z) = \mathcal{C} [(\pm)_{\gz}\;2i]^{(\pm)_b \;|m|}~ e^{-(\pm)_{\gz}\;iz}\;z^{1/2(\pm)_b \;|m|}\;.
\eeq
Here we use a subscript to denote which $\pm$ sign we are working with.
It follows that
\beq \label{Eq:Superpotential7}
W_k(k\gr) = -  \frac{d}{d z} \Bigg\{ \ln \left[ e^{-(\pm)_{\gz}\;iz}\;z^{1/2 (\pm)_b \;|m|}~ \mathfrak{F}(\tfrac{1}{2} (\pm)_b\;|m|,1 (\pm)_b\;2|m|,(\pm)_{\gz} 2iz)\right]\Bigg\}\;,
\eeq
and
\beq \label{Eq:RadWfcn1}
\sqrt{\gr} \psi_{k,\rho,(\pm)_b\;|m|}(\gr) \propto e^{-(\pm)_{\gz}\;iz}\;z^{1/2 (\pm)_b \;|m|}~ \mathfrak{F}(\tfrac{1}{2}\;(\pm)_b\;|m|,1\;(\pm)_b\;2|m|,(\pm)_{\gz} 2iz)\;.
\eeq

For case \# 1, we use Eqs.~\eqref{Eq:Mtilde} and \eqref{Eq:RadWfcn1} to write  
\beq
\psi_{k,\rho,-|m|}(\gr) \propto e^{\mp iz} z^{|m|}\;M(\tfrac{1}{2}+|m|,1+2|m|,\pm 2iz)\;,\;z=k\gr\;.
\eeq
But (See 13.6.9 of DLMF and 13.6.1 of AS.),
\beq
J_{|m|} (z) = \frac{1}{\Gamma(|m|+1)}e^{\mp iz} \left(\frac{z}{2}\right)^{|m|} M(|m|+\tfrac{1}{2},2|m|+1,\pm 2iz) ,
\eeq
where $J$ denotes the Bessel function.  (The $\mp$ and $\pm$ are consistent because $a, b$, and $z$ are real.)  Consequently, since 
\beq
J_{-m}(z) = (-1)^m J_{m}(z) ,
\eeq
(See 10.4.1 of DLMF and 9.1.5 of AS.) for $m \ge 0$, we can take
\beq \label{Eq:P-}
\psi_{k,\rho,-|m|}(\gr) \propto J_{-|m|} (k\gr)
\eeq
and
\beq \label{Eq:Solution2a}
\Psi_{k,-|m|} ({\gr, \phi }) = \mathcal{N}_{k,-|m|} J_{-|m|} (k\gr) e^{-i|m|\phi} , \;  E_k = \frac{\hbar^2 k^2}{2M}\;,
\eeq
where $\mathcal{N}_{k,-|m|}$ is the normalization constant.  Obviously the corresponding superpotential is real.  Thus this case gives the usual solution for $m \le 0$. 

For case \# 3, Eq.\eqref{Eq:RadWfcn1} allows us to write
\beq
\psi_{k,\rho,|m|}(\gr) \propto e^{\mp iz} z^{|m|}\;M(\tfrac{1}{2}+|m|,1+2|m|,\pm 2iz)\;,\;z=k\gr\;,
\eeq
so that
\beq \label{Eq:P+}
\psi_{k,\rho,|m|}(\gr) \propto J_{|m|} (k\gr)
\eeq
and
\beq \label{Eq:Solution2b}
\Psi_{k,|m|} ({\gr, \phi }) = \mathcal{N}_{k,|m|} J_{|m|} (k\gr) e^{i|m|\phi} , \;  E_k = \frac{\hbar^2 k^2}{2M}\;,
\eeq
where $\mathcal{N}_{k,|m|}$ is the normalization constant.  Obviously the corresponding superpotential is also real.  Thus this case gives the usual solution for $m \ge 0$.

For case \# 2 and case \# 4,
\beq \label{Eq:RadWfcn2}
\psi_{k,\rho,\mp |m|}(\gr) \propto e^{\mp iz} z^{\mp |m|} U(\tfrac{1}{2}\mp |m|,1\mp 2|m|,\pm 2iz)\;.
\eeq
According to 13.2.19 of DLMF and 13.5.9 of AS,
\begin{align} \label{Eq:U for b=1}
U(a,1,\gz) = &- \frac{1}{\Gamma(a)} \left[\; \ln \gz + \Psi (a) + 2 \gamma \;\right] + 0(\gz\ln \gz)\;,\;\text{as}\;\gz \to 0 ,  \\ &\text{where}\;\Psi (x) = \Gamma'(x)/\Gamma (x) ,\;\text{and}\; \gamma\;\text{is Euler's constant} \notag \;.
\end{align}
Thus $\psi_{k,\rho,0}(\gr)$ diverges as $\gr \to 0$.  According to 13.2.22 of DLMF and 13.5.12 of AS,
\beq \label{Eq:U for b<0}
U(a,b,\gz) = \frac{\Gamma (1-b)}{\Gamma (1+a-b)} + 0(\gz)\;,\;\text{as}\;\gz \to 0\;,\;\text{for}\;\Re(b) \le 0, b \ne 0\;.
\eeq
Thus $\psi_{k,\rho,-|m|}(\gr)$ with $m>0$ diverges as $\gr \to 0$.  According to 13.2.16 of DLMF (and in disagreement with 13.5.6 of AS),
\beq \label{Eq:U with b>2}
U(a,b,\gz) = \frac{\Gamma(b-1)}{\Gamma(a)} \gz^{1-b} + 0(\gz^{2-\Re(b)})\;,\; \text{as}\;\gz \to 0, \;\text{for}\;\Re(b) \ge 2, b \ne 2\;.
\eeq 
Thus $\psi_{k,\rho,|m|}(\gr)$ with $m>0$ diverges as $\gr \to 0$.  This means item \#2 and item \#4 can not yield an acceptable solution, where the wavefunction is finite everywhere.

We see that in Eqs.~\eqref{Eq:Solution2a} and \eqref{Eq:Solution2b}, that we have a complete solution to the problem, but in this case we did need two different cases to generate all of the solutions.  Moreover, we needed only a single factorization.  No factorization chain of auxiliary Hamiltonians is required. Instead, we simply vary $k$ to get the different solutions. Trying to construct a factorization chain now fails, because the zeros of the wavefunction (singularities of the superpotentials) do not match between the original Hamiltonian and the auxiliary Hamiltonians. This does not allow the system to have a general solution for the factorization chain and is the reason why we must use the single-shot factorization approach.

\section{Free Particle in Three Dimensions} \label{S:3D}

For a free particle in three-dimensions, we use spherical coordinates and factor the effective radial Hamiltonian given by Eqs. \eqref{Eq:VT} and \eqref{eq:H}.  Again, the simplest choice for $\gz$ is $\gz = cz$. Moreover, because of the form of $V_{\text{eff}} (\h{r})$, we also take $z = kr$. Equation \eqref{Eq:Zeta} then reduces to
\beq
(c^2 + 4) + \frac{2(2a-b)c}{z} + \frac{b(b-2)- 4l(l+1)}{z^2} = 0 ,
\eeq
where we have again used Eq. \eqref{Eq:Energy 1}.  We see that we must now take
\beq
c=\pm 2i, b=2a, b=-2l\;\text{or}\;2l+2\;.
\eeq
Note that for $b=-2l$\;, we must use $\wta$ and $U$, and for $b=2l+2$ , we must use $M$ and $U$.  It follows that
\begin{align}
c &=\pm2i, a = -l, b = -2l\;,~\text{allowed for}~\wta~\text{and}~U~, \\
c &=\pm2i, a = l +1, b = 2l + 2\;,~\text{allowed for}~M~\text{and}~U~.
\end{align}
So we have the following table of possibilities.  
\begin{table}[hbt] 
\begin{tabular}[t]{||l|l|l|l|l|l|l||}
\hline \hline
Case\#&$\;a$      &$\;b$          &$\quad c$   &\;$\mathfrak{F}\;$&\;Comments            \\ \hline \hline
1      &$\;-l\;$&$\;-2l\;$   &$\;\pm 2i\;$&$\;\wta \phantom{\Big |}$         &\;Usual Solution                 \\ \hline
2      &$\;-l\;$&$\;-2l\;$   &$\;\pm 2i\;$&$\;\,U\phantom{\Big |}$          &$\;r\psi_r(r) \not\to 0$ as $r \to 0$ \\  \hline
3      &$\;l+1$ &$\;2(l+1)\;$&$\;\pm 2i\;$&$\;M\phantom{\Big |}$            &\;Usual Solutions                \\ \hline
4      &$\;l+1$ &$\;2(l+1)$  &$\;\pm 2i\;$&$\;\,U\phantom{\Big |}$          &$\;r\psi_r(r) \not\to 0$ as $r \to 0$ \\ \hline \hline
\end{tabular}
\caption{Possible solutions for the free particle in spherical coordinates.} \label{T:Table3}
\end{table} 

From Eq.\eqref{Eq:h1}, we obtain
\beq \label{Eq:f4}
h(z) = \mathcal{C} (\pm 2i)^{(b-1)/2}\;e^{\mp iz}\;z^{b/2}\;.
\eeq

Equations \eqref{Eq:Superpotential2} and \eqref{Eq:f4} then yield the superpotential as
\beq \label{Eq:Superpotential7a}
W_k(kx) = -  \frac{d}{d z} \Bigg\{ \ln \left[e^{\mp i z} z^{b/2}\;\mathfrak{F}\left(\tfrac{b}{2},b,\pm2iz\right)\right]\Bigg\}\;,\;\text{with}\;b=-2l\;\text{or}\;2l +2\;,\; z = kr .
\eeq
so that
\beq \label{Eq:R}
r\psi_{kr}(r) \propto  e^{\mp iz} z^{b/2} \mathfrak{F}\left(\tfrac{b}{2},b,\pm 2iz \right)\;,\;\text{with}\;b=- 2 l\;\text{or}\;2l+2\;,\;z=kr\;.
\eeq

From the table, Eq.~\eqref{Eq:Mtilde}, and Eq.~\eqref{Eq:R}, we see that for item \#1, the radial wavefunctions satisfy
\beq
\psi_{k,r,l} (r) \propto e^{\mp i z} z^{l} M(l + 1,2l + 2,\pm 2iz), z=kr .
\eeq
We have (See 13.6.9 of DLMF and 13.6.4 of AS.)
\beq
M(l + 1,2l + 2,\pm 2iz) = \Gamma (\tfrac{3}{2}+l)e^{\pm iz}\left(\frac{2}{z}\right)^{l + 1/2} J_{l+1/2} (z) . 
\eeq
(The $\pm$'s are allowed because $a,~b$, and $z$ are real.)  Moreover, the spherical Bessel function is defined by (See 10.47.3 of DLMF and 10.1.1 of AS.)
\beq
j_{l} (z) = \sqrt{\frac{\pi}{2z}} J_{l + 1/2} (z).
\eeq 
Thus
\beq
\psi_{k,r,l} (r) \propto j_{l} (kr).
\eeq
So, we have for this case
\beq \label{Eq:Wavefunction 3}
\Psi _{k,l,m} (r,\theta,\phi) = \mathcal{N}_{k,l,m} j_{l} (kr) Y_{l}^m (\theta,\phi),\;E_k = \frac{\hbar^2k^2}{2M},
\eeq 
where $\mathcal{N}_{k,l,m}$ is a normalization constant and $Y_{l}^m$ is the standard spherical harmonic.  The corresponding superpotential is, of course, real.

For case \#2, we have to work with $U(-l,-2l,\gz)$.  According to 13.2.7 of DLMF,
\beq \label{Eq:U with a an NPI}
U(-m,b,\gz)= (-1)^m \sum_{s=0}^m \binom{m}{s}(b+s)_{m-s}(-\gz)^s\;,
\eeq
where here $m$ is just a non-negative integer.  It thus follows that 
\beq
r\psi_{kr}(r) \sim e^{\mp ikr} r^{-l}\;,\;\text{as}\;r \to 0\;
\eeq
which is not acceptable.  Thus case \#2 does not yield a solution.

For case \#3, we obtain exactly the same result as for case \#1.

For case \#4,
\beq
r\psi_{k,r,l} (r) \propto e^{\mp i z} z^{l +1} U(l + 1,2l + 2,\pm 2iz) .
\eeq
According to  13.2.17 of DLMF and 13.5.7 of AS,
\beq
U(a,2,\gz) = \frac{1}{\gga (a)} \gz^{-1} + 0(\ln \gz),\;\text{for}\;\gz \to 0.
\eeq
and according to Eq.~\eqref{Eq:U with b>2}, $r\psi_{k,r,l}(r)$ diverges as $r \to 0$.  Thus case \#4 does not yield a solution.

We see that in Eqs. \eqref{Eq:Wavefunction 3}, we have a complete solution.  Again, we have used only a single factorization and no factorization chain is needed.
In fact, any attempt to generate a factorization chain fails for the same reason as for the two-dimensional case.

\section{Linear Potential}

For the linear potential, 
\beq
V(x) = Cx,\;\text{with}\;C > 0,
\eeq
we take 
\beq
z = k_0(x - \tfrac{E_k}{C}),\;\text{where}\; k_0 = \left(\tfrac{2MC}{\hbar^2}\right)^{1/3}.
\eeq
Be careful not to conflate $k_0$, derived from the potential, with $k$, which determines the energy eigenvalue.
The simplest reasonable choice for $\gz$ is 
\beq
\gz = c z^d,\;c\;\text{and}\;d\;\text{constants to be determined}\;.
\eeq
Equation \eqref{Eq:Zeta} then reduces to
\beq
c^2 d^2 z^{2d-2} + 2(2a-b) c d^2 z^{d-2} + \frac{d^2[ b(b-2)+1]-1}{z^2} = 4 z \;.
\eeq
We are required to take
\beq
d = \frac{3}{2}, c = \pm \frac{4}{3}, b = 2a,\;\text{and}\;b = \frac{1}{3}\;\text{or}\;\frac{5}{3}\;.
\eeq
We see that we have eight possibilities as shown in the table below.  Since $a$ and $b$ are not integers, we can use $M$ and $U$, or $M$ and $\wta$, or $\wta$ and $U$; we elect the first possibility.  We emphasize that the wavefunctions for case 1 and case 3, case 2 and case 4, case 5 and case 7, and case 6 and case 8, must be linearly independent of one another. 

\begin{table}[htb]
\resizebox{0.73\textwidth}{!}{
\centering
\begin{tabular}{||l|l|l|l|l|l|l|l||}
\hline \hline
Case \#&$\;a$  &$\;b$  &$\;c$    &$\;d$&$\mathfrak{F}\;$&\;Comments \\ \hline \hline
1      &\;1/6\;&\;1/3\;&\;\;4/3\;&\;3/2&$M\;$           &$\;\psi(x)$ Diverges Exponentially as $x \to + \infty$ \\ \hline
2      &\;1/6  &\;1/3  &\;-4/3   &\;3/2&$M\;$           &$\;\psi(x)$ Diverges Exponentially as $x \to + \infty$ \\ \hline
3      &\;1/6  &\;1/3  &\;\;4/3  &\;3/2&$U\;$           &\;Usual Solution \\ \hline
4      &\;1/6  &\;1/3  &\;-4/3   &\;3/2&$U\;$           &$\;\psi(x)$ Diverges Exponentially as $x \to + \infty$ \\ \hline
5      &\;5/6  &\;5/3  &\;\;4/3  &\;3/2&$M\;$          &$\;\psi(x)$ Diverges Exponentially as $x \to + \infty$ \\ \hline
6      &\;5/6  &\;5/3  &\;-4/3   &\;3/2&$M\;$          &$\;\psi(x)$ Diverges Exponentially as $x \to + \infty$ \\ \hline
7      &\;5/6  &\;5/3  &\;\;4/3  &\;3/2&$U\;$          &\;Usual Solution     \\ \hline
8      &\;5/6  &\;5/3  &\;-4/3   &\;3/2&$U\;$          &$\;\psi(x)$ Diverges Exponentially as $x \to + \infty$ \\ \hline \hline
\end{tabular}
}
\caption{This table shows the putative solutions for the linear potential.}
\label{T:Items}
\end{table} 

From Eq.~\eqref{Eq:h1}, with $d = 3/2$, we readily obtain
\beq
h(z) = \mathcal{C} \sqrt{\frac{2}{3}}c^{a-1/2} e^{-\gz/2} \left(\frac{\gz}{c}\right)^{a-1/6} .
\eeq
It follows that
\beq
W_k (k_0x) = - \frac{d}{d z} \Bigg\{ \ln \left[e^{-\gz/2} \left(\frac{\gz}{c}\right)^{a-1/6} \mathfrak{F}(a,2a,\gz)\right]\Bigg\}
\eeq
and
\beq \label{Eq:Wavefunction3}
\psi_k(x) \propto e^{-\gz/2} \left(\frac{\gz}{c}\right)^{a-1/6} \mathfrak{F}(a,2a,\gz) ,
\eeq
where $\mathfrak{F}(a,2a,\gz)$ is either $M(a,2a,\gz)$ or $U(a,2a,\gz)$.

We immediately note that for case 7, we have (See 13.6.11 of DLMF and 13.6.25 of AS.)
\beq \label{Eq:Case7}
U\left(\tfrac{5}{6},\tfrac{5}{3},\tfrac{4}{3}z^{3/2}\right) = \frac{\sqrt{\pi}\; 3^{5/6}}{2^{2/3}}\frac{e^{\tfrac{2}{3}z^{3/2}}}{z} Ai(z) ,
\eeq
and thus
\beq \label{Eq:Wavefunction2}
\psi_k (x) = \mathcal{N} Ai(z), \; z = \left( \frac{2MC}{\hbar^2} \right)^{1/3} \left(x - \frac{E_k}{C}\right),
\eeq
where $Ai(z)$ is the Airy function of the first kind,  $\mathcal{N}$ is the normalization constant, and $E_k=\tfrac{\hbar^2k^2}{2M}$.  We know that $Ai(z) \to 0$ very rapidly as $z \to + \infty$ and oscillates as $z \to -\infty$.  This is precisely the behavior we expect for the wavefunction of a particle moving in the potential $V(x) = Cx$ with $C > 0$.
We see that we have a complete solution for the problem, which is the usual solution, and aside from normalization, the system wavefunction is as given by Eq. \eqref{Eq:Wavefunction2}.

One can readily show that (See 9.7.5 of DLMF, 10.4.59 of AS, and Eq.~(b.4) of Landau and Lifshitz.\cite{Landau-Lifshitz})
\beq
\psi_k (x) \sim \mathcal{N}\frac{1}{2 \sqrt{\pi}} \frac{e^{-\tfrac{2}{3} z^{3/2}}}{z^{1/4} }, \; \text{for} \; x \to + \infty,
\eeq
and (See 9.7.9 of DLMF, 10.4.60 of AS, and Eq.(b.5) of Landau and Lifshitz.\cite{Landau-Lifshitz})
\beq
\psi_k (x) \sim \mathcal{N}\;\frac{1}{\sqrt{\pi}} \frac{1}{|z|^{1/4}} \sin \left( \frac{2}{3} |z|^{3/2} + \frac{\pi}{4} \right) \; \text{for}\;x \to - \infty ,
\eeq
and (See \S 24 of Landau and Lifshitz.\cite{Landau-Lifshitz})
\beq
\int_{-\infty}^{\infty}d x \; \psi_k (x) \psi_{k'} (x) = \delta (E-E'),
\eeq
with
\beq
\mathcal{N} = \frac{(2M)^{1/3}}{ C^{1/6} \hbar^{2/3}} .
\eeq

We have the identity,
\beq \label{Eq:Ub}
U(a,2a,\gz)=e^{\gz/2}\frac{\gz^{1/2-a}}{\sqrt{\pi}}K_{a-1/2}(\gz/2)\;.
\eeq
(See 13.6.10 of DLMF and 13.6.21 of AS.)  Here $K_{\nu}$ is a modified Bessel function.  Accordingly,
\beq
U(\tfrac{1}{6},\tfrac{1}{3},\gz) = \frac{1}{\sqrt{\pi}} e^{\gz/2} \gz^{1/3} K_{1/3} (\gz/2) ,
\eeq
where we have noted that $K_{-\nu} = K_{\nu}$ (See 10.27.3 of DLMF and 9.6.6 of AS.).  But, 
\beq
Ai(z) = \frac{1}{\pi} \sqrt{\frac{z}{3}} K_{1/3} (\gz/2).
\eeq 
(See 9.6.1 of DLMF and 10.4.14 of AS.)  Thus
\beq
U(\tfrac{1}{6},\tfrac{1}{3},\gz) = 2^{2/3}\;3^{1/6}\;\sqrt{\pi}\;e^{\gz/2} Ai(z) .
\eeq
Consequently,
\beq
h(z)\;U(\tfrac{1}{6},\tfrac{1}{3},\gz) \propto 2^{2/3}\;3^{1/6}\;\sqrt{\pi}\;Ai(z) .
\eeq
Thus case 3 also yields the usual solution.

From Table \ref{T:Items} and Eq.~\eqref{Eq:Wavefunction3}, we have for item 1 and item 5
\beq
\psi_k(x) \propto e^{-\gz/2} \gz^{a-1/6} M(a,2a,\gz)\;, a = \{\tfrac{1}{6},\tfrac{5}{6}\},\;\gz = + \tfrac{4}{3} z^{3/2}\;.
\eeq 
We know that, according to 13.2.4 and 13.7.2 of DLMF, provided $a \ne$ a non-positive integer,
\begin{align} \label{Eq:Asymptotic M}
M(a,b,\gz) \sim& \left[ \frac{\Gamma (b)}{\Gamma (a)} e^{\gz} \gz^{a-b} + \frac{\Gamma(b)}{\Gamma(b-a)}e^{\pm i\pi a}\gz^{-a}\right] \left[1+0(\gz^{-1}\right]
\;,\;\text{as}\;|\gz| \to \infty\;,\; \\  \text{for}&\;- \frac{\pi}{2} + \delta \le \pm  \arg \gz \le \frac{3 \pi}{2} - \delta \notag ,
\end{align} 
where $\delta$ is an arbitrarily small positive constant.  This is mostly as given in 13.1.4, 13.1.5, and 13.5.1 of AS, and on page 60 of Slater~\cite{Slater}, although it appears to disagree with 13.2.23 of DLMF.  Nevertheless, it is certainly true that for $a \ne$ a non-positive integer,
\beq
\psi_k (x) \sim e^{\gz/2}\gz^{-1/6},\;\text{as}\;\gz \to + \infty .
\eeq
Thus for item 1 and item 5, where $c = + 4/3$, so that $\gz \to + \infty$ as $z \to + \infty$, $\psi(x) \to \infty$ exponentially as $x \to + \infty$, and so we do not have an acceptable solution.

For item 2 and item 6, for which $c = - 4/3$, $\gz \to - \infty$ as $z \to + \infty$.  Thus we use Eqs.~\eqref{Eq:Wavefunction3} and \eqref{Eq:Asymptotic M} to write
\beq
\psi_k (x) \sim e^{-\gz/2}\gz^{-1/6},\;\text{as}\;\gz \to - \infty .
\eeq
We see that $\psi_k(x) \to \infty$ exponentially as $x \to + \infty$, which precludes a solution for item 2 and item 6.

Finally, we turn to item 4 and item 8, where $\ga = -4/3$ and $F(a,b,\gz) = U(a,b,\gz)$.  We know that in general (See DLMF 13.2.6.)
\beq
U(a,b,\gz) \sim \gz^{-a}\;,\;\text{as}\;|\gz| \to \infty .
\eeq
It follows that
\beq
\psi_k (x) \sim e^{-\gz/2} \gz^{-1/6}\;,\;\text{as}\;|\gz| \to \infty .
\eeq
Thus, since $\gz \to - \infty$ as $z \to + \infty$ for item 4 and item 8, it follows that $\psi_k(x) \to \infty$ exponentially as $x \to + \infty$, and so these cases do not yield a solution.

We see that two of the eight cases, case 3 and case 7, yield the usual solution.  We also note that in contrast to the free particle problems, it is $U$, the Tricomi function, that yields the solutions.

\section{Continuum of the Hydrogenic Atom}

We have already solved the hydrogen problem heuristically. Now, we revisit the problem to show how to solve it using the more general approach, which does not use any additional ansatzes and which follows our general methodology for solving these types of problems.
   
  Since there are no additive terms to $r$ that appear in $V_{\text{eff}}(r)$, we set $\ga$ to $0$ so that $z = kr$.  The simplest choice for $\gz$ that appears to have a chance of working is simply $\gz = cz$ , where $c$ is a constant to be determined.  Equation \eqref{Eq:Zeta} then reduces to
\beq
c^2\left[1 + \frac{2(2a-b)}{cz} + \frac{b(b-2)}{c^2z^2}\right] = \frac{8M}{\hbar^2 k^2}\left[\frac{l(l+1)\hbar^2}{2Mr^2} - \frac{Ze^2}{r} - E\right] ,
\eeq
and
\beq
(c^2 + 4) + 2\left[(2a-b)c + \frac{4Z}{k\tilde{a}_0}\right]\frac{1}{z} + \frac{b(b-2) - 4l(l+1)}{z^2} = 0\;,
\eeq
where
\beq
\tilde{a}_0 = \frac{\hbar^2}{Me^2}
\eeq
and
\beq
E_k = \frac{\hbar^2 k^2}{2M} \; .
\eeq
Note that since $M$ is the reduced mass of the atom, not merely the electron mass, $\tilde{a}_0$ is the reduced Bohr radius, as introduced previously.  This requires
\beq \label{Eq:Parameters10}
c = \pm 2 i,\;a = \frac{b}{2} \pm i\;\frac{Z}{k\tilde{a}_0},\;b = 2l +2\;\text{or} - 2l\; .
\eeq
We note that for $b = 2l + 2, \widetilde{M}(a,b,\gz)$ does not exist, and for $b = - 2l, M(a,b,\gz)$ does not exist.  Consequently
\begin{align}
&c = \pm 2 i,\;a = l+1\pm i\;\frac{Z}{k\tilde{a}_0},\;b = 2l + 2\;,\; \text{allows for} \;M\; \text{and}\;U\;,\\
&c = \pm 2 i,\;a = -l \pm i\;\frac{Z}{k\tilde{a}_0},\;b = - 2l\;, \;\text{allows for}\;\wta\; \text{and}\;U\;.
\end{align}
Consequently, we have eight distinct cases to consider, as detailed in the table below.  
\begin{table}[htb]
\begin{tabular}{||l|l|l|l|l|l|l|l||}
\hline \hline
Item \#&\quad a                 &\quad b    &\, c   &$\;\mathfrak{F}\;$&Case&\; Comments \\ \hline \hline
1      &$l+1+iZ/k\tilde{a}_0$&$2(l+1)$& +2i   &$\;M\; \phantom{\Big{|}} $         &1.C&\;Usual Solution \\ \hline
2      &$l+1+iZ/k\tilde{a}_0$&$2(l+1)$& +2i   &$\,\;U\ \phantom{\Big{|}}$         &1.C&\;$\psi_{k,r,0}(r) \to \infty$ as $r\to 0$ \\ \hline
3      &$l+1-iZ/k\tilde{a}_0$&$2(l+1)$&\,\;-2i&$\;M\;  \phantom{\Big{|}}$         &1.C&\;Usual Solution \\ \hline
4      &$l+1-iZ/k\tilde{a}_0$&$2(l+1)$&\,\;-2i&$\,\;U\;\phantom{\Big{|}}$         &1.C&\;$\psi_{k,r,0}(r) \to-\infty$ as $r\to 0$ \\ \hline
5      &$-l+iZ/k\tilde{a}_0$ &$-2l$   & +2i   &$\;\wta\;\phantom{\Big{|}}$        &1.B&\;$\psi_{k,r,l}(r) \to \infty$ as $r\to 0$ \\ \hline
6      &$-l+iZ/k\tilde{a}_0$ &$-2l$   & +2i   &$\,\;U\;\phantom{\Big{|}}$         &1.B&\;$\psi_{k,r,0}(r) \to \infty$ as $r\to 0$ \\ \hline
7      &$-l-iZ/k\tilde{a}_0$ &$-2l$   &\,\;-2i&$\;\wta\ \phantom{\Big{|}}$        &1.B&\;$\psi_{k,r,l}(r) \to \infty$ as $r\to 0$ \\  \hline
8      &$-l-iZ/k\tilde{a}_0$ &$-2l$   &\,\;-2i&$\,\;U\;\phantom{\Big{|}}$         &1.B&\;$\psi_{k,r,0}(r) \to \infty$ as $r\to 0$ \\ \hline \hline
\end{tabular}   \label{T:ContinuumStates}  
\caption{Possible Solutions for the continuum states of the Hydrogenic atom}
\end{table}

Equation \eqref{Eq:h1} with $\gz = cz = \pm2iz$ gives
\beq \label{Eq:h2}
h(z) = \mathcal{C} \frac{e^{-\gz/2} \gz^{b/2}}{c^{1/2}} = \mathcal{C} (\pm2i)^{(b-1)/2} e^{\mp iz} z^{b/2}\;.
\eeq
Equations \eqref{Eq:Superpotential5} and \eqref{Eq:h2} together give for the superpotential
\beq \label{Eq:Superpotential5b}
W_k(kr) = -  \frac{d}{d z} \Bigg\{ \ln \left[e^{\mp i z}z^{b/2}~\mathfrak{F}(a,b,\pm 2iz)\right]\Bigg\} , z = kr\;,
\eeq
and for the radial wavefunction,
\beq \label{Eq:RadWfnc1}
r\psi_{k,r,l}(r) \propto e^{\mp ikr} (kr)^{b/2}~\mathfrak{F}(a,b,\pm2ikr)
\eeq
with $a, b$, and $\mathfrak{F}$ as given in the table.  We note that this superpotential guarantees that we produce the correct Hamiltonian with an effective potential given by
\beq
V_{\text{eff}}(r) = \frac{l(l+1)\hbar^2}{2Mr^2} - \frac{Ze^2}{r} .
\eeq
We shall prove that, for the cases for which it gives a solution, this superpotential is real.  

For item \#1 and item \#3, we see that the continuum radial wavefunction for the hydrogenic atom is given by
\beq \label{Eq:RadWfnc}
\psi_{k,r,l} (r) = \mathcal{N}_{k,l}\;e^{\mp i kr} (kr)^{l} M\left (l+1\pm i\tfrac{Z}{k\tilde{a}_0},2l+2,\pm 2ikr\right) ,
\eeq
where $\mathcal{N}_{k,l}$ is a normalization constant.  The $\pm$ is allowed because $kr, l$, and $Z/k\tilde{a}_0$ are real and it is the usual solution.  (See \S\;36 of Landau and Lifshitz~\cite{Landau-Lifshitz}.)

For item \#5 and item \#7, Eqs.~\eqref{Eq:Mtilde} and \eqref{Eq:RadWfnc1} yield
\beq 
\psi_{k,r,l} (r)\propto e^{\mp i kr} (kr)^{-l-1} M\left (l+1\pm i\tfrac{Z}{k\tilde{a}_0},2l+2,\pm 2ikr\right )\;.
\eeq
which diverges as $r \to 0$, and thus does not yield an acceptable solution.

Now we turn to item \#2 and item \#4, for which
\beq
\psi_{k,r,l}(r) \propto e^{\mp ikr}\;(kr)^{l}\;U\left (l+1\pm i\tfrac{Z}{k\tilde{a}_0},2l+2,\pm 2ikr\right )\;.
\eeq
According to 13.2.16 and 13.2.17 of DLMF,
\beq 
\psi_{k,r,l} (r)\sim e^{\mp i kr} (kr)^{-l-1}\;\text{as}\;r \to 0\;,
\eeq
which diverges as $r \to 0$, and thus also does not yield an acceptable solution. 

Finally, for item \#6 and item \#8, we obtain 
\beq
\psi_{k,r,l}(r) \propto e^{\mp ikr}\;(kr)^{-l-1}\;U\left (-l \pm i\tfrac{Z}{k\tilde{a}_0},-2l,\pm 2ikr\right )\;.
\eeq
We use 13.2.40 of DLMF or 13.1.29 of AS to rewrite this as
\beq
\psi_{k,l,r}(r) \propto e^{\mp ikr}\;(kr)^{l}\;U\left (l +1 \pm i\tfrac{Z}{k\tilde{a}_0},2(l +1),\pm 2ikr\right )\;.
\eeq
This is the same as for item \#2 and item \#4, so that no acceptable solution results.
 
From Eq.\eqref{Eq:RadWfnc}, it follows that
\beq 
\psi_{k,r,l}^* (r) = \mathcal{N}_{k,l}^* \;e^{\pm i kr} (kr)^{l} M\left (l+1\mp i\tfrac{Z}{k\tilde{a}_0},2l+2,\mp 2ikr\right )\;.
\eeq
We use 13.2.39 of DLMF or 13.1.27 of AS to rewrite this as
\beq
\psi_{k,r,l}^* (r) = \mathcal{N}_{k,l}^* \;e^{\mp i kr} (kr)^{l} M\left (l+1\pm i\tfrac{Z}{k\tilde{a}_0},2l+2,\pm 2ikr\right )\;.
\eeq
We see that the superpotential is real and, providing that $\mathcal{N}_{k,l}$ is real, $\psi_{k,r,l}(r)$ is also real.

\section{The One-Dimensional Morse Potential}

We finally consider the one-dimensional Morse potential, given by
\beq
V(x) = D\left(e^{-2k_0x} - 2e^{-k_0x}\right),~- \infty < x < \infty~.
\eeq
We take 
\beq
z = k_0x,~\text{and}~\gz = c e^{-z},~\text{where}~c~\text{is a constant to be determined}. 
\eeq 
We note that one should not confuse the non-negative wavenumber $k_0$ that appears in the potential with the parameter $k$ that appears in the energy, $E_k=\tfrac{\hbar^2k^2}{2M}$.

Equation \eqref{Eq:Zeta} and the above two equations give
\beq \label{Eq:Zeta2}
\gz^2 \left[1 + \frac{2(2a-b)}{\gz} + \frac{b(b-2)}{\gz^2}\right] + 3 - 2 = 4 \xi^2 \left(\frac{\gz^2}{c^2} - 2 \frac{\gz}{c}\right) - 4\eta^2~ .
\eeq
where
\beq
\eta = \sqrt{\frac{2ME_k}{\hbar^2 k_0^2}}=\frac{k}{k_0},\;\xi = \sqrt{\frac{2MD}{\hbar^2 k_0^2}}.
\eeq
From Eq.~\eqref{Eq:Zeta2}, we immediately see that
\beq
2(2a-b)=-~\frac{8}{c}~\xi^2,~ b(b-2)+1=-4~\eta^2,~ 1=\frac{4}{c^2}~\xi^2~.
\eeq
Thus 
\beq \label{Eq:parameters5}
a = \frac{1}{2}-(\pm)_c~\xi+(\pm)_b~i\eta,~b = 1+(\pm)_b~2i\eta~,~c = (\pm)_c~2 \xi~.
\eeq
Given that $a \not \in \mbi^{\le 0}$ and $b \not \in \mbi$ (except possibly for $E_k=0$, but more on that later) , we can take $\mathfrak{F}$ to be $M$ and $U$.  We thus have the table of eight possibilities shown below.  The Case designations are taken from the table in Ref. \cite{Properties}. 
\begin{table}[!hbp]
\begin{tabular}{||l|l|l|l|l|l|l|l||}
\hline \hline
Item \#&\quad a                 &\quad b    &\, c   &$\;\mathfrak{F}\;  $&Case&\; Comments \\ \hline \hline
1      &$\tfrac{1}{2}-\xi +i\eta$&$1+2i\eta$&$+2\xi$&$\;M\;\phantom{\Big{|}}  $&1.A&\;$\psi(x)\to \infty$~\text{as~}$x\to -\infty$ \\ \hline
2      &$\tfrac{1}{2}-\xi +i\eta$&$1+2i\eta$&$+2\xi$&$\,\;U\ \phantom{\Big{|}}$&1.A&\;Usual Solution \\ \hline
3      &$\tfrac{1}{2}+\xi +i\eta$&$1+2i\eta$&$-2\xi$&$\;M\; \phantom{\Big{|}} $&1.A&\;$\psi(x)\to \infty$~\text{as~}$x\to -\infty$ \\ \hline
4      &$\tfrac{1}{2}+\xi +i\eta$&$1+2i\eta$&$-2\xi$&$\,\;U\;\phantom{\Big{|}}$&1.A&\;$\psi(x)\to \infty$~\text{as~}$x\to -\infty$ \\ \hline
5      &$\tfrac{1}{2}-\xi -i\eta$&$1-2i\eta$&$+2\xi$&$\;M\; \phantom{\Big{|}} $&1.A&\;$\psi(x)\to \infty$~\text{as~}$x\to -\infty$ \\ \hline
6      &$\tfrac{1}{2}-\xi-i\eta$&$1-2i\eta$&$+2\xi$&\;$\;U\;\phantom{\Big{|}}$&1.A&\;Usual Solution \\ \hline
7      &$\tfrac{1}{2}+\xi-i\eta$&$1-2i\eta$&$-2\xi$&$\;M\; \phantom{\Big{|}} $&1.A&\;$\psi(x)\to \infty$~\text{as~}$x\to -\infty$ \\  \hline
8      &$\tfrac{1}{2}+\xi-i\eta$&$1-2i\eta$&$-2\xi$&$\,\;U\;\phantom{\Big{|}}$&1.A&\;$\psi(x)\to \infty$~\text{as~}$x\to -\infty$ \\ \hline \hline
\end{tabular}   
\caption{Possible Solutions for the Continuum States of the One-Dimensional or Cartesian Morse Potential}
\end{table} \label{T:MContinuumStates}  

From Eqs.\eqref{Eq:h1} and \eqref{Eq:parameters5}, and $\gz=ce^{-z}$, we have 
\beq
h(z) = \mp~i~\mathcal{C}~ e^{-\gz/2} \gz^{(b-1)/2} = \mp~i~\mathcal{C}~ e^{-\gz/2} \gz^{[(\pm)_b] ~ i \eta}~.
\eeq
We note that
\beq
\gz \to 0~\text{as}~z \to + \infty, \gz \to (\pm)_c ~ \infty~\text{as}~z \to - \infty~. 
\eeq
Moreover (See DLMF 13.2.4 and 13.7.2 and AS 13.5.1.), 
\begin{align} \label{Eq:limM2}
M(a,b,\gz) \sim &\left[ \frac{\gga (b)}{\gga (a)} e^{\gz} \gz^{a-b} + \frac{\gga(b)}{\gga(b-a)}e^{\pm i\pi a}\gz^{-a}\right] \left[1+\mathcal{O}(\gz^{-1})\right]
\;,\;\text{as}\;|\gz| \to \infty\;,\; \\  &\text{for}\;- \frac{\pi}{2} < \pm  \arg(z) < \frac{3 \pi}{2} ,\;\text{unless}\;a \in \mathbf{Z}^{\le0}\; \text{or}\;b-a \in \mathbf{Z}^{\le0}\;. \notag
\end{align}
Thus, for $\mathfrak{F} = M$ and $\gz = +2~\xi~e^{-z}$,
\beq
h(z)~M(a,b,\gz) = \mp~i~\mathcal{C}~e^{-\gz/2} \gz^{(b-1)/2}M(a,b,\gz) \sim e^{\gz/2} \gz^{a-(b+1)/{2}}~\text{as}~z \to - \infty \notag
\eeq	
and
\beq
h(z)~M(a,b,\gz) = \sim e^{\gz/2} \gz^{-\tfrac{1}{2}-\xi}~\to \infty~\text{as}~z \to - \infty~.
\eeq
For $\mathfrak{F} = M$ and $\gz = -2~\xi~e^{-z}$,
\beq
h(z)~M(a,b,\gz) = \mp~i~\mathcal{C}~e^{-\gz/2} \gz^{(b-1)/2}M(a,b,\gz) \sim e^{-\gz/2} \gz^{-a+(b-1)/{2}}~\text{as}~z \to - \infty \notag
\eeq	
and
\beq
h(z)~M(a,b,\gz) = \sim e^{-\gz/2} \gz^{-\tfrac{1}{2}-\xi}~\to \infty~\text{as}~z \to - \infty~.
\eeq
Consequently, we cannot use $M(a,b,\gz)$~at all and items \# 1, 3, 5, and 7 do not yield a solution.  

We see that we are left with $\mathfrak{F} = U$.  We note that (See DLMF 13.2.6 and AS 13.5.2.  Note that our constraint on the phase is consistent with a branch cut along $(-\infty,0]$.)
\beq	
U(a,b,\gz) \sim \gz^{-a}~,~\text{as}~|\gz| \to \infty~,~\text{for}~|\arg \gz| < \pi~. 
\eeq
Then
\beq
h(z)~U(a,b,\gz) = \mp~i~\mathcal{C}~e^{-\gz/2} \gz^{(b-1)/2}U(a,b,\gz) \sim e^{-\gz/2} \gz^{-a+(b-1)/{2}}~\text{as}~z \to - \infty \notag
\eeq	
and
\beq
h(z)~U(a,b,\gz) = \sim e^{-\gz/2} \gz^{-\tfrac{1}{2}+(\pm)_c~\xi}~\to 
\begin{cases}
	0 \\
	\infty
\end{cases}
~\text{as}~z \to - \infty~.
\eeq
Thus items \# 4 and 8 do not yield a solution.  Moreover (See DLMF 13.2.18,
\beq
U(a,b,\gz) = \frac{\gga(b-1)}{\gga(a)} \gz^{1-b} + \frac{\gga(1-b)}{\gga(1+a-b)} + \mathcal{O}(\gz^{2-\Re(b)})\;,\;\text{as}\;\gz \to 0, 
\eeq
for $1 \le \Re(b) < 2$, $b \ne 1$,
so that $\psi_k (x)$ is well behaved for $x \to + \infty$.

Consequently, we are left with items \# 2 and 6, and we have
\beq \label{Eq:Superpotential 10}
W_k(k_0x) = - \frac{d}{d z}\ln \left[\gz^{\pm i \eta} e^{-\gz/2} U\left(\tfrac{1}{2} -\xi\pm i \eta ,1\pm i 2 \eta,\gz \right)\right],~\gz = 2\xi e^{-k_0x},
\eeq
and
\beq \label{Eq:MorseWavefunction}
\psi_k(x) = \mathcal{N}_k~\gz^{\pm i \eta} e^{-\gz/2} U\left(\tfrac{1}{2} - \xi \pm i \eta,1\pm i 2 \eta,\gz \right)~,~\gz = + 2\xi e^{-k_0x},
\eeq
where, of course $\mathcal{N}_k$ is the normalization constant.  
We have obtained a complete solution for all energy eigenstates with $E>0$ with just a single factorization.

Given that $\mathcal{N}_k$ is real,
\beq
\psi_k(x)^* = \mathcal{N}_k~\gz^{\mp i \eta} e^{-\gz/2} U\left(\tfrac{1}{2} - \xi \mp i \eta,1\mp i 2 \eta,\gz \right)~,
\eeq 
The Kummer transformation for U is (See 13.2.40 of DLMF and 13.1.29  AS.)
\beq
U(a,b,\gz) = \gz^{1-b} U(1+a-b,2-b,\gz)~.
\eeq
From Eq. \eqref{Eq:parameters5} with $(\pm)_c = +1$, we have
\beq
1+a-b=1+\frac{1}{2}-\xi \pm i\eta -1 \mp i 2 \eta = \frac{1}{2}-\xi\mp i \eta~, 2-b=1\mp 2i\eta~.
\eeq
thus
\beq
\psi_k(x) = \mathcal{N}_k~\gz^{\pm i \eta} e^{-\gz/2}~\gz^{\mp 2i\eta} U\left(\tfrac{1}{2} - \xi \mp i \eta,1\mp i 2 \eta,\gz \right) ,
\eeq
and
\beq
\psi_k (x) = \psi_k (x)^* ~.
\eeq
Of course, the superpotential is also real,
\beq
W_k (k_0x)^* = W_k (k_0x)~.
\eeq
 
We note that
\beq
E_k=0 \Longrightarrow a = \frac{1}{2}-(\pm)_c ~ \xi~,~b=1~,~c=\pm2 ~ \xi~,~\eta = 0~.
\eeq
If $a \not \in \mbi^{\le 0}~,~b=1$, then $U \not \propto M$ and $\wt = M(a,b,\gz)$, so that we can still take $\mathfrak{F} = M$ and $U$.  Moreover, we can't use $M$ at all and $U$ with $c = - 2 \xi$ for the same reasons as before.  We thus obtain
\beq \label{Eq:Morse0Wavefunction}
\psi_0 (x) = \mathcal{N}_0 ~e^{-\gz/2}~U\left(\tfrac{1}{2}-\xi, 1, \gz\right)~, 
\eeq
which is our previous result, as given by Eq. \eqref{Eq:MorseWavefunction}, with $E=0$~and $\eta = 0$.

We also note that
\beq
E_k=0,~\gz=2\xi e^{-k_0x},~a = - m,~\text{with}~m \in \mbi^{\ge 0} \Longrightarrow \xi = m+\frac{1}{2}~,
\eeq
so that $a \in \mbi^{\le 0}$ is possible.  Then $U \propto M$, and since $b=1 \Longrightarrow \wta = M$, it follows that $M, \wta$, and $U \Longrightarrow$ only one solution.  We are in cell 4.C of the table in Ref. \cite{Properties}, and so the second solution is given by DLMF 13.2.8.  With $a = - m, n=0, k \to s$, we have
\begin{align}
\mathfrak{F}(-m,1,\gz) =  -& \sum_{s\,=\,0}^m \frac{(-m)_s}{(1)_s}\;\frac{\gz^s}{s!}\;[\;\ln(\gz) + \psi(1+m-s) - 2\psi(1+s)\;] + \notag \\ +& \;(-1)^{1+m}\;m!\;\sum_{s\,=\,1+m}^{\infty} \frac{(s-1-m)!}{(1)_s}\;\frac{\gz^s}{s!} \;\notag \;,
\end{align}
where here $\psi(v) = \frac{\Gamma'(v)}{\Gamma(v)}$ is the digamma function.  Since $(1)_s = s!$, we see that 
\beq
\mathfrak{F}(-m, 1, \gz) \sim e^{\gz}~\text{as}~|\gz| \to \infty~,
\eeq
and thus is not an acceptable solution.  Consequently Eq. \eqref{Eq:Morse0Wavefunction} is valid even when $\tfrac{1}{2}-\xi \in \mbi^{\le 0}$.

There is one final point.  The wavefunction given by \eqref{Eq:MorseWavefunction} is not what is found in the literature!  The resolution of this is straightforward.  For $E_k\ne 0, b$~is not an integer, and so we can use Eqs. \eqref{Eq:Mtilde}, \eqref{Eq:MorseWavefunction}, and (See 13.2.42 of DLMF, 13.1.3 of AS, and \S 1.3 of Slater.\cite{Slater})
\beq \label{Eq:U1}
U(a,b,\gz) = \frac{\gga(1-b)}{\gga(1+a-b)}\;M(a,b,\gz) + \frac{\gga(b-1)}{\gga(a)}\;\widetilde{M}(a,b,\gz)\;.
\eeq
to write
\beq
\psi_k (x) = \mathcal{N}_k \gz^{\pm i\eta} e^{-\gz/2} \Bigg[ \frac{\gga(1-b)}{\gga(1+a-b)}\;M(a,b,\gz) + \frac{\gga(b-1)}{\gga(a)}\;\gz^{1-b}~M(1+a-b,2-b,\gz)\Bigg]\;.
\eeq
We have
\beq
a = \frac{1}{2} - \xi \pm i\eta ,~b = 1 \pm i 2 \eta ,~1+a-b = \frac{1}{2}-\xi \mp i \eta ,~2-b = 1\mp i 2 \eta ~.
\eeq
Accordingly,
\begin{align}
\psi_k (x) = &\mathcal{N}_k  e^{-\gz/2} \Bigg[ \gz^{\pm i \eta} \frac{\gga(\mp 2i\eta)}{\gga(\tfrac{1}{2}-\xi \mp i\eta)}\;M(\tfrac{1}{2}-\xi \pm i\eta,1\pm 2i\eta,\gz)~+ \\ &+~\gz^{\mp i\eta}~\frac{\gga(\pm 2i\eta)}{\gga(\tfrac{1}{2}-\xi \pm i\eta )}~M(\tfrac{1}{2}-\xi \mp i \eta,1 \mp 2i\eta ,\gz)\Bigg]\;.
\end{align}
This is the form found in the literature \cite{Nicholls} and that has also recently been obtained with the Laplace method for solving the Schr\"odinger equation  \cite{Canfield}. Indeed, the equality of these two forms is one of the reasons why working with confluent hypergeometric functions can be tricky.
 
\section{Discussion and Conclusions} 

We have developed the formalism of what we term ``single-shot factorization'' and used it to solve the continuum problems for the free particle in one, two, and three dimensions, the linear potential in one dimension, the hydrogenic atom, and the one-dimensional Morse potential.  In all cases, we have used only a single factorization (that depends on a parameter) and produced the different solutions by varying the parameter.  Hence, a factorization chain is not required.  This is in contrast to the standard factorization treatment of bound-state problems, where a chain of auxiliary Hamiltonians is employed to obtain all of the eigenvalues and eigenstates of the original Hamiltonian. Note that for the one-dimensional free particle, a factorization chain is possible, but it does not produce all of the allowed energy eigenstates. It occurs due to the uniform spacing of the zeros of the wavefunction.
In summary, we have demonstrated that the standard lore that factorization can be used only for bound states and not for continuum states is simply not true.

The single-shot factorization method employs two steps. In the first step, the function $\zeta(z)$ is determined along with a number of parameters that go into the solutions. This is done independently of knowing the wavefunction for the energy eigenstate. Then, the appropriate  confluent-hypergeometric function is chosen, by enforcing the requirements of the wavefunction, This is a straightforward approach, but it requires working carefully with the asymptotic behavior of confluent hypergeometric functions and carefully working with the two possible linearly independent solutions of the confluent hypergeometric differential equation.

From what we have done, it appears that the single-shot factorization method should apply to bound states in addition to continuum states.  We have found that to be the case and have successfully applied the single-shot factorization method to the one-dimensional simple harmonic oscillator, the bound states of hydrogen and the one-dimensional Morse potential \cite{Mathews1}.  Moreover, we expect that Eqs. \eqref{Eq:Zeta} and \eqref{Eq:h1} - \eqref{Eq:psiofq} encompass many of the known solutions of the single-particle Schr\"odinger equation and we expect that the single-shot factorization method can be extended to include all known solutions of the single-particle Schr\"odinger equation.  Specifically, there are solutions of the Schr\"odinger equation that cannot be expressed in terms of \chf, but require full hypergeometric functions.  We are in the process of extending the procedure used in this paper to deal with those cases.  We also anticipate that Eq. \eqref{Eq:Gen Cond} may reveal some new solutions.  These are all active areas of investigation. 

From the work presented here, we have deduced a general factorization procedure for solving continuum and bound state problems for which the wavefunctions can be expressed in terms of \chf.   Our results are summarized by Eqs.~\eqref{Eq:VT}, \eqref{eq:H}, \eqref{Eq:Wavefunctions}, \eqref{Eq:Superpotential2}, \eqref{Eq:Mtilde},  \eqref{Eq:Gen Cond}, \eqref{Eq:Zeta}, \eqref{Eq:g2}, \eqref{Eq:f1}, \eqref{Eq:f2}, \eqref{Eq:h1}, \eqref{Eq:psiofq}, and \eqref{Eq:Superpotential5}. The approach yields a factorization of the Hamiltonian, with a positive energy, that is independent of the parameter, and hence leads to all of the continuum solutions by simply varying the parameter. It can also be viewed as a systematic approach to solving the Ricatti equation for the superpotential, for cases solvable with confluent hypergeometric functions.

The fact that the general procedure of single-shot factorization is identical for the three types of \chf\;when we are pursuing the simpler cases stemming from Eqs.~\eqref{Eq:g 1} and \eqref{Eq:Zeta}, rather than Eq.~\eqref{Eq:Gen Cond}, means that we need not decide which of the three types of \chf \; to use until very near the end of the calculation, when the limiting behavior of the putative wavefunctions becomes apparent.  This is noteworthy because all three types of \chf \; give rise to Bessel functions, Hermite functions, Laguerre functions, and Weber or parabolic cylinder functions. These solutions all appear in continuum solutions in quantum mechanics. Finally, note that this work directly finds the continuum solution rather than the more common analytic continuation approaches used in most textbooks. We feel this direct approach is much more concrete and understandable, once one has mastered the relevant results needed about the confluent hypergeometric functions.

\vspace{6pt} 




\funding{This research was funded by the National Science Foundation under Grant No.~PHY-1620555 and by the McDevitt bequest at Georgetown University. }

\acknowledgments{We acknowledge useful discussions with Mark Esrick and ZuYao Teoh.}

\authorcontributions{Conceptualization, J.~K.~F.; Methodology, J.~K.~F.~ and W.~N.~M.;  Formal Analysis, W.~N.~M.;  Writing – Original Draft Preparation, W.~N.~M.; Writing – Review and Editing, J.~K.~F.;  Funding Acquisition, J.~K.~F.}

\dataavailability{There is no data used in this study.} 

\conflictsofinterest{The authors declare no conflicts of interest.}

\end{paracol}
\reftitle{References}

\end{document}